\documentclass[%
 reprint,
 pre,
 floatfix,
 superscriptaddress
]{revtex4-2}
\usepackage{pgf}
\usepackage{amsmath}
\usepackage{amssymb}
\usepackage{mathtools}
\usepackage{graphicx}
\usepackage{color}
\usepackage{array}
\usepackage{adjustbox}
\usepackage{placeins}
\usepackage{hyperref}
\usepackage{ifthen}
\newcolumntype{L}{>{$}l<{$}} 

\setcitestyle{numbers,square,sort&compress}

\newcommand{\subfigimg}[3][,]{%
  \setbox1=\hbox{\includegraphics[#1]{#3}}
  \leavevmode\rlap{\usebox1}
  \rlap{\hspace*{10pt}\raisebox{\dimexpr\ht1-2\baselineskip}{#2}}
  \phantom{\usebox1}
}

\begin{document}

\title{Signatures of the interplay between chaos and local criticality \\ on the dynamics of scrambling in many-body systems}

\author{Felix Meier}
\affiliation{University of Duisburg-Essen, Lotharstr. 1, 47048 Duisburg, Germany}
\author{Mathias Steinhuber}
\affiliation{University of Regensburg, Universitätsstr. 31, 93040 Regensburg, Germany}
\author{Juan Diego Urbina}
\affiliation{University of Regensburg, Universitätsstr. 31, 93040 Regensburg, Germany}
\author{Daniel Waltner}
\affiliation{University of Duisburg-Essen, Lotharstr. 1, 47048 Duisburg, Germany}
\author{Thomas Guhr}
\affiliation{University of Duisburg-Essen, Lotharstr. 1, 47048 Duisburg, Germany}

\begin{abstract}
 
Fast scrambling, quantified by the exponential initial growth of Out-of-Time-Ordered Correlators (OTOCs), is the ability to efficiently spread quantum correlations among the degrees of freedom of interacting systems, and constitutes a characteristic signature of local unstable dynamics. As such, it may equally manifest both in systems displaying chaos or in integrable systems around criticality. Here, we go beyond these extreme regimes with an exhaustive study of the interplay between local criticality and chaos right at the intricate phase space region where the  integrability-chaos transition first appears. We address systems with a well defined classical (mean-field) limit, as coupled large spins and Bose-Hubbard chains, thus allowing for semiclassical analysis. Our aim is to investigate the dependence of the exponential growth of the OTOCs, defining the quantum Lyapunov exponent $\lambda_{\textrm{q}}$ on quantities derived from the classical system with mixed phase space, specifically the local stability exponent of a fixed point $\lambda_{\textrm{loc}}$ as well as the maximal Lyapunov exponent $\lambda_{\textrm{L}}$ of the chaotic region around it. By extensive numerical simulations covering a wide range of parameters we give support to a conjectured linear dependence $2\lambda_{\textrm{q}}=a\lambda_{\textrm{L}}+b\lambda_{\textrm{loc}}$, providing a simple route to characterize scrambling at the border between chaos and integrability.
\end{abstract}


\maketitle

\section{Introduction}
\label{ch1}

The temporal growth properties of Out-of-Time-Ordered Correlators (OTOCs), first introduced in \cite{OldschoolOTOC}, provide a useful tool in the understanding of scrambling of quantum correlations, the ubiquitous mechanism behind many important many-body emergent phenomena, from equilibration/thermalization \cite{JD1_1,JD1_2,JD1_3,JD1_4,JD1_5} to the information loss paradox and black hole physics  \cite{JD2_1,Maldacena}, and ultimatively to the quantum signatures of chaos \cite{JD3_1}.

In the context of the quantum-classical correspondence and the signatures of quantum chaos, by means of the methodology of canonical quantization OTOCs  can be shown to have a classical limit related to the Poisson brackets of the classical system. In the semiclassical limit, this in turn relates their growth behavior to quantifiers of dynamical instability and chaoticity \cite{Maldacena,OldschoolOTOC}.

A reason for why the OTOCs have gained considerable attention for studies of this type is that it offers a way to make statements on the thermodynamical high particle limit of a system \cite{Maldacena} while still maintaining its semiclassical interpretation, while quantum chaos traditionally has been an area focused on systems with few degrees of freedom as long as there ought to be performed a semiclassical limit explicitly (see however \cite{JD4_1}).

In this work we are only considering systems with low dimensional classical analogs and investigate in detail how the quantity $\lambda_{\textrm{q}}$ characterizing the OTOCs for a time regime shorter than the Eherenfest time, in which the OTOC grows exponentially as $C(t)\sim \exp(2\lambda_{\rm q} t)$, is related to classical quantities characterising the classical dynamics for a system with mixed phase space.

In previous works \cite{CoupledTopsPhysRevE60,CoupledTopsPhysRevE67} it was shown that the entanglement entropy, another quantity related to investigations of the quantum-classical correspondence with dynamically chaotic classical limit, in some cases grows exponentially with an exponent depending on the sum of classical Lyapunov exponents of subsystems.
Furthermore in \cite{LyapForRegularClassLimit,InvertedOsc} it was shown that exponential growth of an OTOC can be attributed to a local instability - a hyperbolic fixed point - even in an otherwise regular system, that in a many-body context is a standard indicator of critical behavior \cite{JD5_1,JD5_2}.

For a system and regimes with classical mixed phase space, hence, where both quantities $\lambda_{\text{loc}},~\lambda_L$ can be present at the same time we propose the hypothesis that in its exponential regime, an OTOC grows with an exponent depending both on the local as well as global stability classifiers, namely the stability exponent of a fixed point as well as the Lyapunov exponent of the phase-space region in which the fixed point is located. 
For systems with at least locally unstable classical limit it is assumed that an OTOC grows polynomially until a time $t_{\rm 0}$ referred to as the dissipation or ergodic time, then enters the exponential regime that we are interested in until a time $t_{\rm E}$ given by the Ehrenfest time \cite{Ehrenfest1927,BERMAN1978,chirikov1981dynamical,danielBook} and finally saturates to a constant value depending on the dimension of the quantum mechanical state space \cite{Maldacena,TypicalOTOCGrowthMBL,WeakQMChaos,OTOCAfterScramblingTime}.
In works like \cite{Maldacena,WeakQMChaos} this time $t_{\rm E}$ is also referred to as the scrambling time $t_\ast$, which is a synonymous choice of terminology.
Writing from the perspective of semiclassics we refer to it as the Ehrenfest time that marks the breakdown of the quantum to classical correspondence, rather than as the scrambling time which is the name used when analyzing the time it takes for the scrambling of information about initial states setting in as an effect of the temporal dynamics.
Further notable recent works on the topics related to the growth behavior of OTOCs are \cite{Newref1,Newref3,Newref4,Newref5,Newref6,Newref7,Newref8,Newref9,Newref10,LiebRobinsonBound,OTOCChaosIntegrableOperatorFront, new_more_otoc, new_more_otoc2, new_more_otoc3}.
Related experimental results can be found e.g. in \cite{new_experimental, new_experimental2, new_experimental3}.

In Sec.~\ref{ch2} we present the basic theoretical notions starting from the type of OTOC used, its relation to classical quantities and the formulation of the main hypothesis that we test in this work.
Afterwards in Sec.~\ref{ch3} we apply the theory on a spin system similar to the one defined in~\cite{Maram} as well as compare the results with a similar investigation of a Bose-Hubbard system in Sec.~\ref{ch3_2}.
Finally Sec.~\ref{ch4} contains the conclusions we can draw from the obtained results.

\section{Exponential Growth of OTOCs}
\subsection{Definitions and Concepts}
\label{ch2}
For our purposes we use as the OTOC of choice the commutator squared, defined as
\begin{align}
	\label{DefinitionOTOC}
	&C^{\hat{A}\hat{B}}_\rho (t):= \left\langle \big|\big|\big[\hat{A}(t),\hat{B}(0)\big]\big|\big|^2\right\rangle_\rho, \\
	&\rho:=|z\rangle\langle z|,\,\,\, z\in \mathcal{P},\,\,\, ||\hat{\mathcal{O}}||^2 := \hat{\mathcal{O}}\hat{\mathcal{O}}^\dagger,\nonumber
\end{align}
where $|z\rangle$ is a coherent state situated at a point $z$ of a symplectic manifold $\mathcal{P}$ that is serving as the classical phase space of the Hamiltonian system associated to the quantum Hamiltonian generating the time evolution $U(t)$ and with the Heisenberg time evolution
\begin{align}
	\hat{A}(t) := U^\dagger(t)\hat{A}U(t)\,\,\,\text{.}
\end{align}

Here, the particular form of the OTOC follows the standard literature \cite{Maldacena}. To gain intuition about its key characteristic features, we can compare Eq.~\eqref{DefinitionOTOC}
against the susceptibilities in the framework of linear response theory~\cite{toda2012statistical} that are defined in exactly the same way but without the square operation and are deeply connected with the notion of causality. The presence of the square in the definition of the OTOC has then two effects. First, it makes impossible to time-order it in a causal form, making it very different from a susceptibility. Second, OTOCs are then defined as expectations of positive-definite objects, and therefore produce a large signal even under averaging, making them quite robust objects.

By means of the quantum-classical correspondence principle, up to the Ehrenfest time $t_{\rm E}$ we can associate the squared commutator with a squared Poisson bracket and the quantum time evolution $\hat{A}(t)=\alpha_t(\hat{A}):=U^\dagger(t)\hat{A}U(t)$ with the classical flow given by a symplectomorphism $\phi_t$ as $\phi_t(z_0):=z(t)$ and finally obtain
\begin{align}
	\label{OTOC2Poisson}
	&C^{\hat{A}\hat{B}}_\rho(t) \rightarrow \left(\{A\circ \phi_t,B\}_{(z)}\right)^2,\\
	& A,B\in\mathcal{C}^\infty(\mathcal{P},\mathbb{R}),\,\,\,\phi_t \in \mathrm{Symp}(\mathcal{P},\omega)\subset\mathcal{C}^\infty(\mathcal{P},\mathcal{P}).\nonumber
\end{align}
Here $\mathrm{Symp}(\mathcal{P},\omega)$ are the symplectomorphisms~\cite{Waldmann,Marsden} of $\mathcal{P}$ relative to the symplectic form $\omega$ and $\mathcal{C}^\infty(\mathcal{P},\mathcal{P})$ are the smooth functions from $\mathcal{P}$ to itself.
While we always have to work in a chart representation when performing the numerics showcased in the later chapters, we remark that neither of the spaces containing the dynamics occurring in this work is identical to an Euclidean space, so that it can not be identified globally with a chart domain \cite{Nakahara}.
Instead of writing down all chart maps explicitly we employ slight abuse of notation by denoting a point $x\in\mathcal{P}$ implicitly as its chart representative as $x=z(x)=(z^1,\cdots,z^{2f})=(q^1,\cdots,q^f,p_1,\cdots,p_f)$ so that $z$ is to be seen as either an abstract point in $\mathcal{P}$, a tuple of chart maps or a tuple of coordinates depending on where it appears in an equation.
The quantities $A,B$ without hats are the classical observables corresponding to the operators $\hat{A},\hat{B}$ and $\{\,,\,\}_{(z)}$ is the Poisson bracket evaluated at a point $z\in\mathcal{P}$, defined by means of the symplectic form $\omega$ in terms of Hamiltonian vector fields~$X^f,~X^g$~\cite{Waldmann,Marsden}
\begin{align}
	\label{definition_Poisson}
	&\{f,g\}_{(z)}:=\omega_{(z)}(X^f_{(z)},X^g_{(z)})\\
	&X^f:=(df)^{\#_\omega},\,\,\, f,g\in\mathcal{C}^\infty(\mathcal{P},\mathbb{R})\,\,\,\text{.}\nonumber
\end{align}
Here $\#_\omega$ is the musical isomorphism \cite{Waldmann} induced by the skew-bilinear form $\omega$ that translates a co-vector $df$ into a vector $X^f$.
We denote the differential of a function $f:\mathcal{P}\rightarrow\mathcal{P}$ as $df$ which is a linear map $df:\mathrm{T}_z\mathcal{P}\rightarrow\mathrm{T}_{f(z)}\mathcal{P}$ generalizing the concept of a derivative to the setting of smooth manifolds.
With respect to coordinates $z=(z^\mu)$ the differential corresponds to the Jacobian $[\partial \tilde{f}^\mu(z)/\partial z^\nu]$ of the chart-representative $\tilde{f}=z\circ f\circ z^{-1}$.
If we choose the observables $A,B$ as coordinate functions $z^{\mu_A}, z^{\mu_B}$ corresponding to the $\mu_A$-th and $\mu_B$-th coordinates from the chart functions $z=(q^1,\cdots,q^f,p_1,\cdots,p_f)$ that form a system of canonical coordinates, we can identify the Poisson bracket with a component of the linearized flow:
\begin{align}
	\label{Poisson2linFlow}
	\{z^{\mu_A}\circ\phi_t,z^{\mu_B}\}_{(z)}=\omega^{\mu_B \nu} \left[d{\phi_t}_{(z)}\right]^{\mu_A}_{\,\,\,\nu}\propto \exp(\lambda_{\text{loc}}(z))\,\,\,\text{.}
\end{align}
Here $\omega^{\mu\nu}$ are the matrix components of the inverse of the matrix representation of the symplectic form $\omega$, $d{\phi_t}_{(z)}$ is the linearized flow at $z$ and in the whole expression the Einstein summation of equal upper and lower indices is implied.
A derivation of this expression can be found in App.~\ref{appendix1}.
For each region $U\subseteq\mathcal{P}$ that is an invariant region with respect to the flow, $\phi_t(U)\subseteq U$, a concept of dynamical stability can be established.
The most basic invariant subsets of phase space are period $t$ periodic orbits $\mathcal{PO}(\phi_t,t)$ with respect to the map that is given by the flow $\phi_t$ among which the simplest case is that of a fixed point.
If we assume $z$ to be a fixed point, meaning a periodic orbit with no minimal period, then the domain and target of the linearized flow are the same, $d{\phi_t}_{(z)}:\mathrm{T}_z\mathcal{P}\rightarrow \mathrm{T}_{\phi_t(z)=z}\mathcal{P}$ and we can meaningfully speak of its eigenvalues~\cite{Froeschle}, which are then given by local stability exponents $\lambda_{\text{loc}}(z)$ as in the previous Eq.~\eqref{Poisson2linFlow}.
We call a direction defined by a tangent vector from the coordinate basis ${\partial/\partial z^\mu}_{(z)}$ at $z$ stable, if its corresponding exponent $[d{\phi_t}_{(z)}]^\mu_{\,\,\,\mu}\propto\lambda_{\text{loc}}(z)$ is purely imaginary, while we call it unstable if the exponent has a positive real part.
In the general case where $z$ is not on a periodic orbit, the concept of eigenvalues for the linearized flow breaks down and the corresponding quantifier of stability becomes the maximal Lyapunov exponent which takes account of the whole region the point is situated in.
The maximal Lyapunov exponent is the maximal element of the Lyapunov spectrum and will from hereon always just be referred to as just the Lyapunov exponent. 
This language is chosen since we do not require the other elements from the Lyapunov spectrum.
The Lyapunov exponent of a regular region in phase space, meaning a region that, up to subsets of measure zero, only contains stable states, evaluates to zero since in non-chaotic systems, nearby trajectories only deviate linearly while the Lyapunov exponent measures the rate of exponential divergence~\cite{Meiss,Froeschle}.
On the other hand a chaotic region, meaning one only comprised of unstable points, is characterised by a positive Lyapunov exponent.
The Lyapunov exponent usually can only be computed numerically e.g. by the approximation~\cite{Meiss,Froeschle}
\begin{align}
	\label{definition_Lyapunov}
	&\lambda_{\rm L}(z_0,V_0)\approx \frac{1}{t}\ln\left(\frac{||\tilde{V}(t)||}{||V_0||}\right),\\
	& \tilde{V}(t):=d{\phi_t}_{(z_0)}(V_0)\in \mathcal{C}^\infty(\mathbb{R},\mathrm{T}_{\phi_t(z_0)}\mathcal{P}),\nonumber
\end{align}
where $||.||$ is a suitable norm, $\tilde{V}(t)$ is a parametrised curve in the tangent bundle $\mathrm{T}\mathcal{P}$ to phase space, obtained by the application of the linearized flow to the starting vector $V_0$ and $t$ is a time that has to be taken as large as possible to improve the approximation.
Since Hamiltonian mechanics does not prescribe such a norm on phase space on physical grounds, it has to be seen rather as a mathematical ambiguity at this point.
However, the physics should not depend on which norm is chosen in particular, such that we can simply pick the Euclidean one.
For compact phase spaces the Lyapunov exponent is independent of the choice of metric inducing this norm \cite{TwoParticleMethod}. In the cases we study in this work this criterion is satisfied having $\mathcal{P}=S^2\times S^2$ for the spin system, while for the Bose-Hubbard system the phase space is $\mathcal{P}=\mathbb{R}^{2L}$, but the dynamics are restricted to the submanifold $S^{2L-1}\subset\mathbb{R}^{2L}$.
Here $V_0\in \mathrm{T}_{z_0}\mathcal{P}$ is a tangent vector, also referred to as a deviation vector, that can be chosen arbitrarily as long as it does not point along a stable manifold of the dynamics.
The region for which the Lyapunov exponent is computed can be chosen by picking an initial state $z_0$ from the same region, but in principle $\lambda_{\rm L}(z_0,V_0)$ does not depend on which state from said region is chosen specifically \cite{Froeschle,Meiss}. 
This underlines the global character of this quantity as opposed to local quantities that depend on explicit points $z\in\mathcal{P}$,
\begin{align}
\label{Lyapunov_CoM_derivation}
 \lambda_{\rm L}(z_0,V_0)=\lambda_{\rm L}({\phi_T}_{(z_0)},{d\phi_T}_{(z_0)}(V_0))\,\,\,\text{.}
\end{align}
Therefore we can see that the Lyapunov exponent only depends on the trajectory of the inserted point $z_0$ and not on the specific point along it.
If the system is ergodic and the point $z_0$ is not a periodic orbit or fixed point, the flow fills up the whole region, on which the Lyapunov exponent is hence constant \cite{Froeschle}.
For an in general mixed phase space one therefore obtains one maximal Lyapunov exponent per invariant subset of phase space, where it is zero if the considered subset is a regular island and positive otherwise.
In this sense, a more logical notation for the Lyapunov exponent of an ergodic region $U\subseteq\mathcal{P}$ would be $\lambda_{\rm L}(U)$ but since the practical computation necessitates the choice of a pair $(z_0\in U,V_0\in \mathrm{T}_{z_0}U)$, we will stick to the previously used notation.

\subsection{Quantum Lyapunov exponents}
For an OTOC we assume three phases of its temporal evolution: 
	\begin{align}
		C^{\hat{A}\hat{B}}_\rho(t)\propto
			\left\{
				\begin{array}{@{\kern2.5pt}lL}
				\hfill P(t)\in\mathrm{Poly}(\mathbb{R}),\,\,\, t\leq t_0\\
				\hfill \exp\left(2\lambda_{\rm q}t\right),\,\,\, t_0 \leq t \leq t_{\rm E}\\
				\text{const} (\mathrm{dim}(\mathcal{H})),\,\,\,  t_{\rm E}\leq t
				\end{array}
			\right. \,\,\,\text{.}
	\end{align}
First it grows polynomially - here $\mathrm{Poly}(\mathbb{R})$ denotes the real valued polynomials - reflecting system components becoming entangled by means of quantum mechanics, secondly it grows exponentially with an exponent $2\lambda_{\rm q}$ as a quantum analogue of classical chaos and finally, after the quantum-classical correspondence breaks down at the Ehrenfest time $t_{\rm E}$ it saturates to a value depending on the dimension of the Hilbert space.
At the borders between two regions there might appear smooth interpolations that can deviate slightly from the behaviors stated~\cite{LyapunovQMSpinChains}.
We refer to the quantity $\lambda_{\rm q}$ as a quantum Lyapunov exponent, which is a quantum mechanical object since it is computed purely from quantum mechanical ingredients.
Its prefactor two appears for the commutator squared as a manifestation of taking a square.
In \cite{Maldacena} it has been argued that for certain regularized OTOCs 

\begin{align}
\label{Definition_regularized_F}
F^{\hat{A}\hat{B},\text{reg}}_\rho (t) := \operatorname{Tr}\left(
    y\hat(A)y\hat{B}(t)y\hat{A}y\hat{B}(t)
\right), \,\,\, y:=\sqrt[4]{\rho}
\end{align}

their large particle limit, which in this sense corresponds to a semiclassical limit when the inverse particle number is taken as $\hbar_{\text{eff}}$, should show exponential growth depending on a classical Lyapunov exponent of a classically corresponding system,
\begin{align}
	\label{Maldacena_hypothesis}
	F^{\hat{A}\hat{B},\text{reg}}_\rho(0)-F^{\hat{A}\hat{B},\text{reg}}_\rho (t) \propto \exp(2\lambda_{\rm L} t)
\end{align}
for some suitable thermal state $\rho$ and times $t\in[1/\lambda_{\rm L},t_{\rm E}= \log (1/\hbar_{\text{eff}} )/\lambda_{\rm L}]$.
Here the term regularization means the splitting up of the quantum state into four parts $y$ as in Eq.~\eqref{Definition_regularized_F} and is only of concern for systems in the context of quantum field theory or in systems in low temperature regimes.
Since the OTOC considered by us can be written in terms of similar but unregularized four point correlators $G^{\hat{A}\hat{B}}_\rho(t), F^{\hat{A}\hat{B}}_\rho(t)$ with various types of operator orderings as \cite{InvertedOsc}
\begin{align}
	\label{OTOC_decomposition}
	& C^{\hat{A}\hat{B}}_\rho (t) = G^{\hat{A}\hat{B}}_\rho(t)+F^{\hat{A}\hat{B}}_\rho(t),\\
    & F^{\hat{A}\hat{B}}_\rho (t) := -\left\langle\hat{A}(t)\hat{B}\hat{A}^{\dagger}(t)\hat{B}^{\dagger}
    +(\hat{A}(t)\hat{B}\hat{A}^{\dagger}(t)\hat{B}^{\dagger})^{\dagger}
    )\right\rangle,\nonumber\\
    & G^{\hat{A}\hat{B}}_\rho (t) := \left\langle\hat{A}(t)\hat{B}\hat{B}^{\dagger}\hat{A}^{\dagger}(t)
    +\hat{B}\hat{A}(t)\hat{A}^{\dagger}(t)\hat{B}^{\dagger}
    \right\rangle\nonumber
\end{align}
we can translate the Maldacena hypothesis \cite{Maldacena} to the realm of the commutator squared OTOC as has been done in \cite{LyapunovQMSpinChains} for example.
In order to reference a particular classical state in the quantum computation, we make use of coherent states instead of a canonical ensemble state.
Furthermore, instead of a classical average over Poisson brackets as in \cite{LyapunovQMSpinChains} that yields a quantity similar to the Lyapunov exponent, we compute the Lyapunov exponent directly, since this aforementioned averaging is related to computing the OTOC with respect to a thermal state.
In \cite{LyapForRegularClassLimit,InvertedOsc} it was shown that even a regular system can have exponentially growing OTOCs where the growth is caused only by a local instability, namely a hyperbolic fixed point on a separatrix.
Furthermore in \cite{CoupledTopsPhysRevE67} it was shown that in certain systems the entanglement entropy semiclassically grows exponentially with the exponent given by a sum of classical stability quantifiers of the classical limits of the subsystems.
Entanglement entropy and Renyi entropy are closely related and the latter has been shown to grow exponentially with a Lyapunov exponent just like the OTOC \cite{OTOCEntropyCorrelation,OTOCEntropyCorrelation2,RenyiEntropLyapunov, new_entropy}.
Based on this fact together with the aforementioned special cases for exponential OTOC growth we hence state the following hypothesis the testing of which is the subject of this work:
\\
\\
\\
\textbf{Hypothesis}\\
For some initial time $t_0$, a coherent state $\rho=|z_0\rangle\langle z_0|$, centered on a fixed point defined as $\phi_t(z_0)=z_0$ for all $t \in\mathbb{R}$ we expect the quantum Lyapunov exponent $\lambda_{\rm q}$ to be given by
\begin{align}
	\label{hypothesis}
	& \lambda_{\rm q}:=\frac{1}{2t}\ln\left(C^{\hat{A}\hat{B}}_\rho(t)\right) \approx f_{\rm q}(\lambda_{\rm L}(z_0+\epsilon,V_0),\lambda_{\text{loc}}(z_0)),\\
	& t\in [t_0,t_{\rm E}],\,\,\,
	 f_{\rm q}\in \mathcal{C}^0(\mathbb{R}^+_0\times \mathbb{R},\mathbb{R}^+_0)\,\,\,\text{,}\nonumber
\end{align}
where $\mathcal{C}^0(\mathbb{R}^+_0\times \mathbb{R},\mathbb{R}^+_0)$ are the positive valued continuous functions on the space in which $(\lambda_{\rm L},\lambda_{\text{loc}})$ is situated.
In \cite{Maldacena} for a purely chaotic classical limit $t_0=1/\lambda_{\rm L}$ is assumed and in \cite{JD5_2} $t_0=1/\lambda_{\text{loc}}$ for a single hyperbolic fixed point within a regular region.
Positivity of the exponent is assumed since $\lambda_{\rm q}$ should describe a growth behavior, not a decrease.
Here we include a small increment $\epsilon$ so that the Lyapunov exponent is computed for the region in which the fixed point $z_0$ lies, instead of using the state $z_0$ itself, which would only result in the local stability exponent again.
The function $f_{\rm q}$ is so far undetermined apart from being continuous because we want $\lambda_{\rm q}$ to depend smoothly on the classical quantities. 
This is because these are themselves continuous functions of the system parameters to be introduced in the application section \ref{ch3}.

\section{Testing the Hypothesis for Interacting Large Spins}
\label{ch3}

We consider a spin system with two coupled spins and Hamiltonian 
\begin{align}
	\label{Definition_spin_hamiltonian}
	&\hat{H} =\\
	&\frac{4J \hat{S}^{(1)}_z \hat{S}^{(2)}_z }{(s+1/2)^2}+
	 \sum\limits_{i=1}^{2}\left(\frac{2\left(b_x \hat{S}^{(i)}_x +  b_y \hat{S}^{(i)}_y + b_z \hat{S}^{(i)}_z \right)}{s+1 /2}\right),\nonumber\\
	&\hbar_{\text{eff}} := \frac{1}{s+1/2}\,\,\,\text{.}\nonumber
\end{align}
Here $J$ is the interaction strength between spins and $(b_x,b_y,b_z)$ are magnetic field components.
The spin operators in the spin $s$ representation for the $i$-th spin are written as $\hat{S}^{(i)}_{\alpha}\in\mathrm{End}(\mathbb{C}^{2s+1}),\,\,\, \alpha\in\{x,y,z,+,-\}$.
The concrete choices of prefactors and definition of $\hbar_{\text{eff}}$ have been influenced by previous works and are motivated in \cite{Maram} where a similar system is considered though with a discretized time evolution.
We remark that we set $\hbar=1$ and that for all considered quantities arbitrary units can be assumed since our results are concerning only the qualitative behavior of the systems studied by us.
In order to make the numerics feasible we have to restrict to two spins, while increasing the spin representation $s$ of the $\mathfrak{su}(2)$ observables $\hat{S}^{(i)}_\alpha$. 
The connection to the classical system is obtained by means of spin coherent states:
\begin{align}
	\label{Definition_spin_coherent_states}
		 &|q,p\rangle = \exp\left(\frac{\arccos(p)}{2}(\hat{S}_+ e^{iq} +\hat{S}_- e^{-iq})\right)|s,s\rangle,\\
		 &|q_1,q_2,p_1,p_2\rangle =|q_1,p_1\rangle\otimes|q_2,p_2\rangle,\nonumber
\end{align}
which are always already connected to a choice of chart.
In this case we use the chart
\begin{align}
    \label{Chart_on_S2xS2}
		 &z: S^2 \times S^2 \rightarrow \mathbb{R}^4\\
		 &\,\,\,\,\,\,\,(n^{(1)}, n^{(2)}) \mapsto (q_1,q_2,p_1,p_2),\nonumber\\
		 &q_{i}:=\arctan\left(\frac{n^{(i)}_y}{n^{(i)}_x}\right),\,\,\, p_i=n^{(i)}_z\,\,\,,\nonumber
\end{align}
on the manifold $S^2 \times S^2$.
Spin coherent states can be understood in terms of the theory of Perelomov coherent states \cite{Stratonovich,Perelomov,GeometryOfQuantumStates}, from which it follows that the symplectic manifold serving as the right classical phase space for an individual spin is given by $(\mathcal{P}_0,\omega_0)=(S^2,\mathrm{vol}_{S^2}\cong \sin(\vartheta)d\vartheta\wedge d\varphi)$.
For the system of two spins we can combine two such state spaces simply as
\begin{align}
    \label{TwoSpinPhasespace}
    (\mathcal{P},\omega)=(\mathcal{P}_0\times\mathcal{P}_0,\omega_0\oplus\omega_0)
\end{align}
to obtain the full phase space $(\mathcal{P},\omega)$.
Here we write the symplectic form that is given by the volume form on the sphere in terms of spherical coordinates, which are related to symplectic coordinates as $(\varphi,\cos(\vartheta))=(q,p)$.
Employing these states to translate spin operators to Bloch vectors, $\langle \hat{S}^{(i)}_\alpha\rangle_\rho = sn^{(i)}_\alpha$, we arrive at the classical Hamiltonian in the $\hbar_{\text{eff}}\rightarrow 0 \Leftrightarrow s\rightarrow \infty$ limit:
\begin{align}
	\label{Classical_hamiltonian}
		& H_i=2\left(b_x\cos(q_i) \sqrt{1-p_i^2} + b_y \sin(q_i) \sqrt{1-p_i^2}+b_z p_i\right)\\
		& H(q_1,q_2,p_1,p_2)= H_1 + H_2 + 4J p_1 p_2\,\,\,\text{.}\nonumber
\end{align}
As a facilitation, in the following we always set the magnetic field $b_y=0$ since the spin system only requires two non-zero magnetic field components to allow for a non-integrable classical limit~\cite{QMSpinChainIntegrability}.
Choosing the operators~$\hat{A}=\hat{S}^{(1)}_z=\hat{B}$ relates the OTOC $C^{\hat{A}\hat{B}}_\rho(t)$ to the following Poisson bracket:
\begin{align}
		&C^{\hat{A}\hat{B}}_\rho(t)=C^{\hat{S}^{(1)}_z\hat{S}^{(1)}_z}_{|z_0\rangle\langle z_0|}(t)
		 \rightarrow \{p_1\circ\phi_t,p_1\}^2_{(z_0)}=\\
		 &=\left(\frac{\partial (p_1\circ \phi_t)(z)}{\partial q_1}|_{z_0}\right)^2=\left(\frac{\partial p_1(t)}{\partial q_1}|_{z_0}\right)^2,\nonumber\\
		 &z_0=({q_0}_1,{q_0}_2,{p_0}_1,{p_0}_2)\,\,\,\text{.}\nonumber
\end{align}
We are numerically looking for hyperbolic fixed points inside of a regular region of the classical system by solving the equation
\begin{align}
	&X^H_{(z)} = {(dH)}_{(z)}^{\#_\omega} = \\
	&=\left[\begin{array}{cccc}
	4J p_2 - \frac{2b_x p_1 \cos(q_1)}{\sqrt{1-p_1^2}} + 2b_z\\
	4J p_1 - \frac{2b_x p_2 \cos(q_2)}{\sqrt{1-p_2^2}} + 2b_z\\
	2 b_x\sqrt{1-p_1^2} \sin(q_1)\\
	2 b_x\sqrt{1-p_2^2} \sin(q_2)
	\end{array}\right]
	\stackrel{\text{!}}{=} 0\nonumber
\end{align}
for the coordinates $z\in\mathbb{R}^4$ while fixing the magnetic field parameters $(b_x,0,b_z)$ and varying the interaction $J$.
For the thus obtained fixed points we compute the Lyapunov exponents of their phase space region by adding a small increment $\epsilon$ to yield a starting point $z+\epsilon$ for employing the previously mentioned method.
For the parameters~$(b_x,b_y,b_z) = (0.05,0,0.05)$ we find the coordinates of a hyperbolic fixed point $z_0\in U\subset z(\mathcal{P})$ within a region with vanishing Lyapunov exponent, $\lambda_{\rm L}(U)=0$.
As we increase the interaction $J$ we can recalculate the coordinates of this fixed point to obtain a curve $J\mapsto z_0(J)$ that smoothly depends on $J$.
Since the dynamics of the system depend on the interaction parameter, the region~$U$ does not stay a regular region for all values of the curve parameter and we can likewise construct a curve of Lyapunov exponents $J\mapsto \lambda_{\rm L}(U,J)$ that will reach positive values for the in general mixed phase space.

\begin{figure*}
\centering
	\includegraphics[width=0.49\textwidth,trim={1 2 3 4},clip=True]{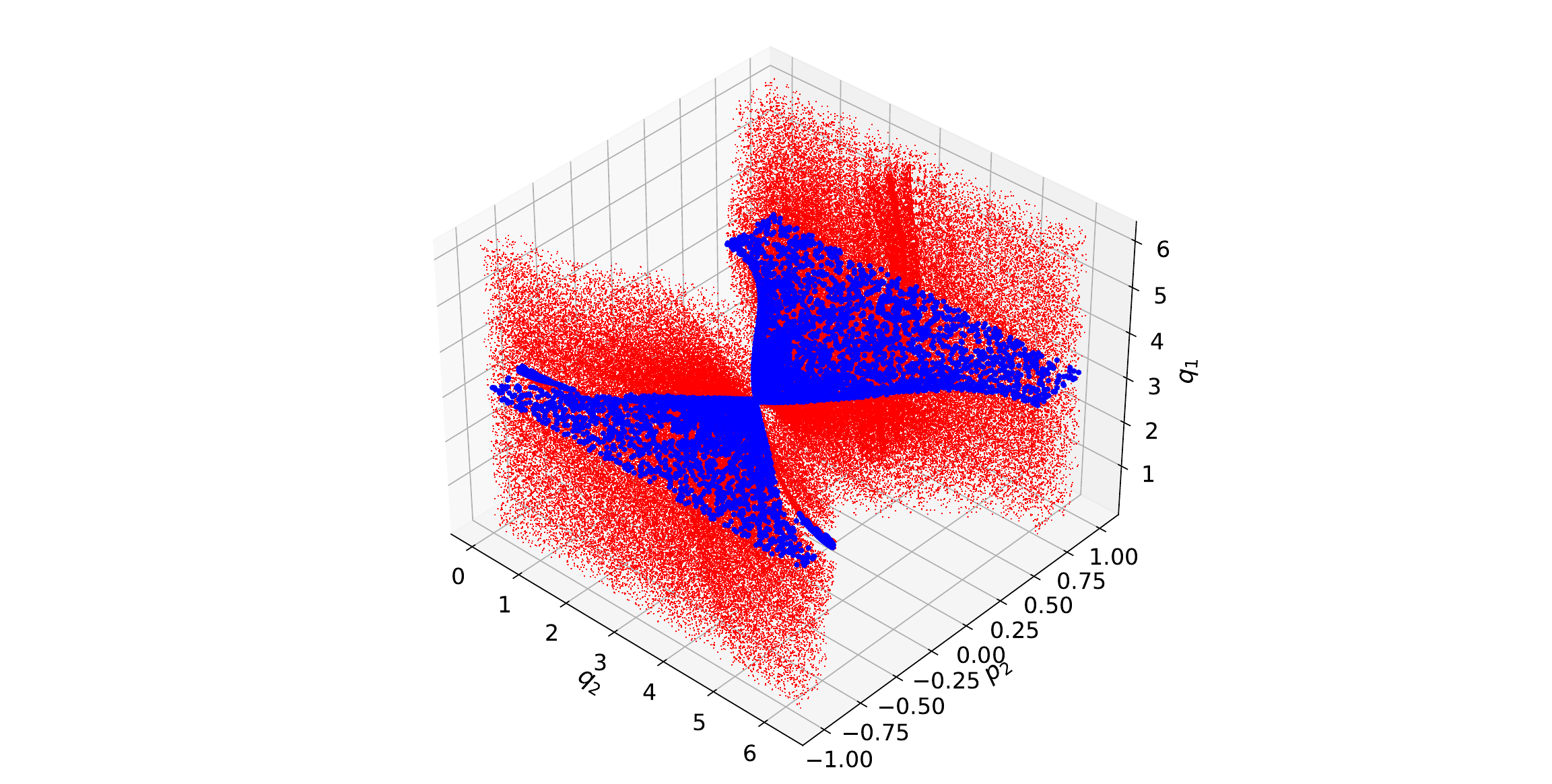}
	\includegraphics[width=0.49\textwidth]{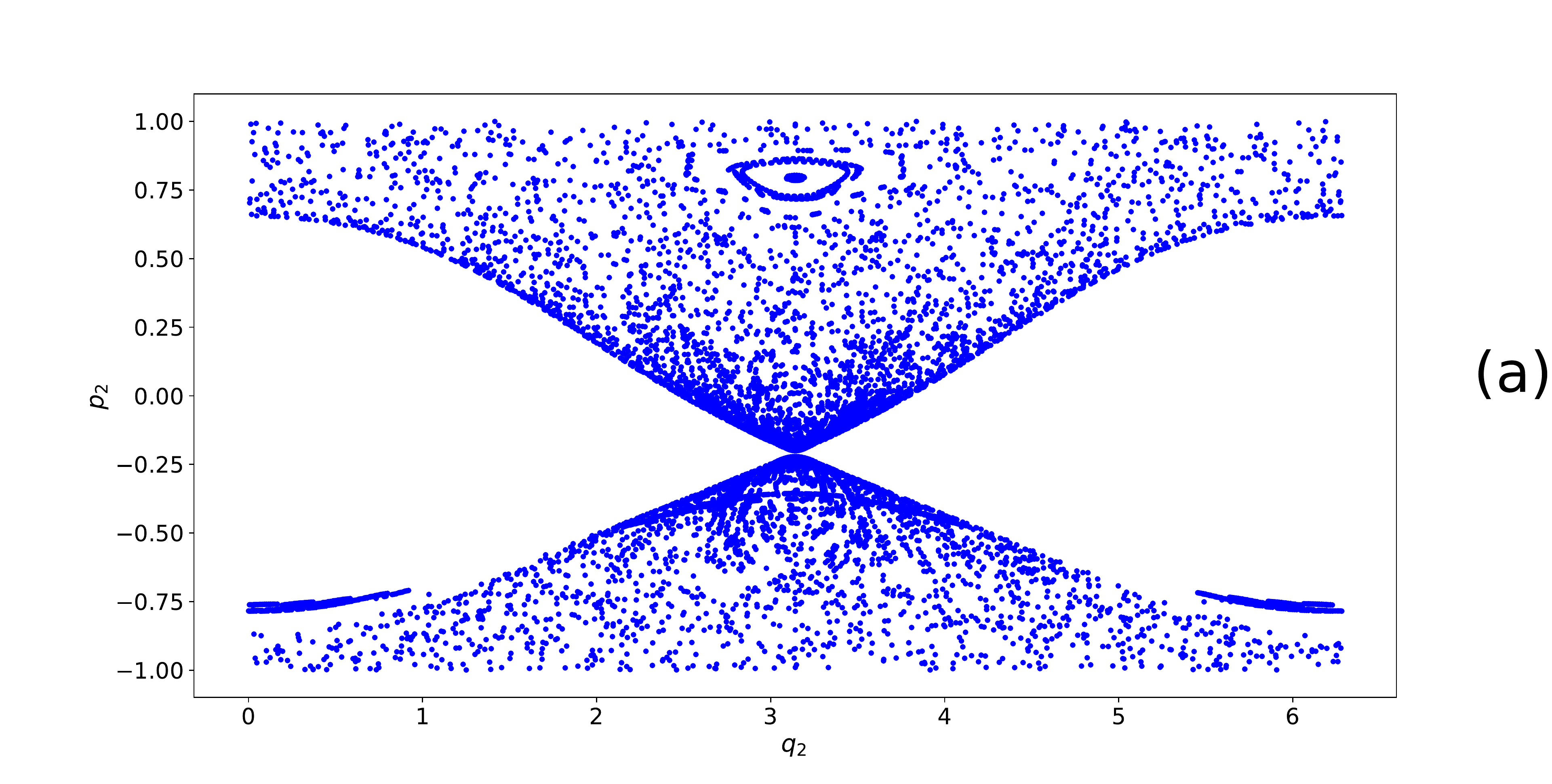}
	\hfill
	\includegraphics[width=0.49\textwidth]{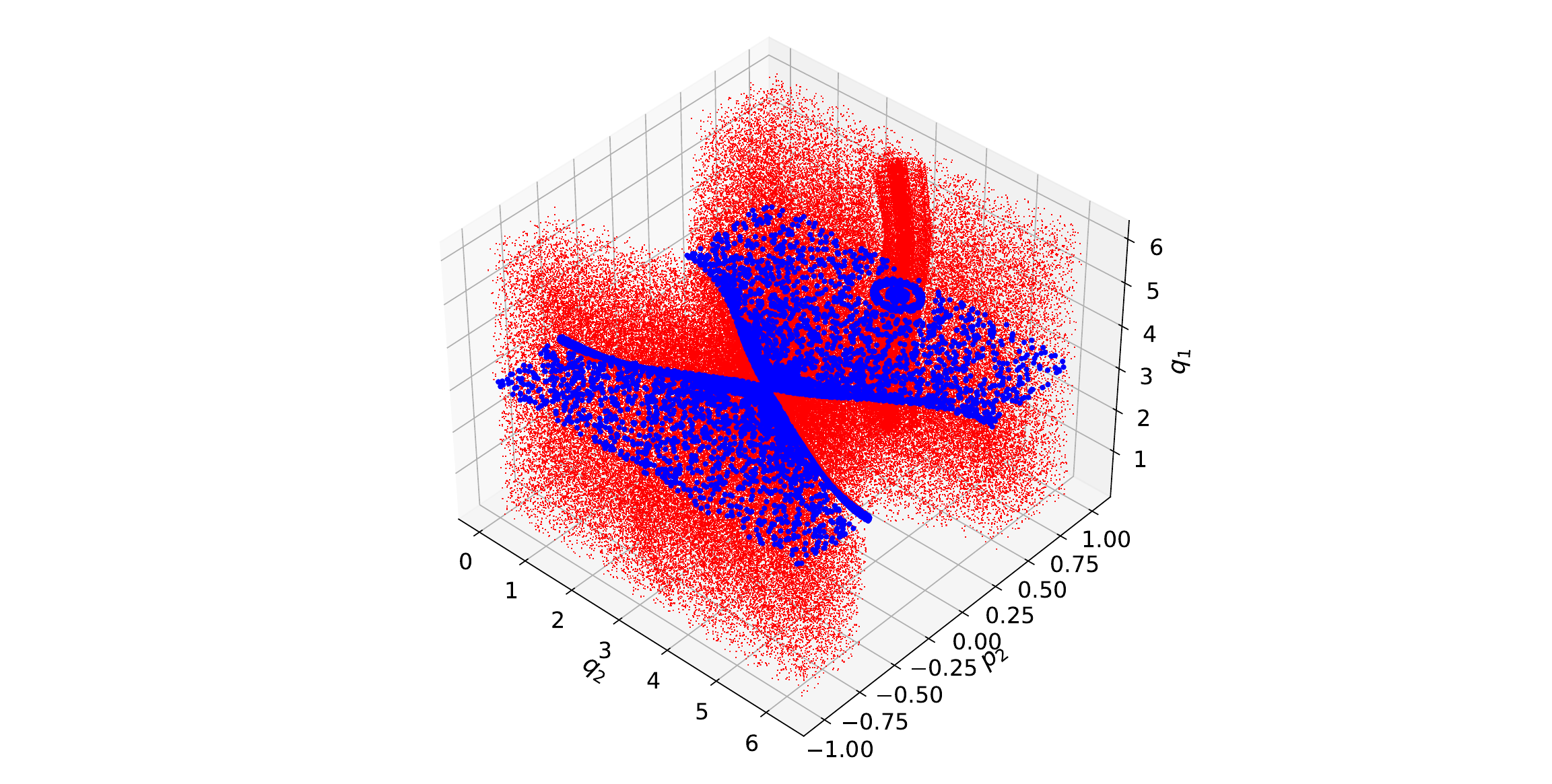}
	\includegraphics[width=0.49\textwidth]{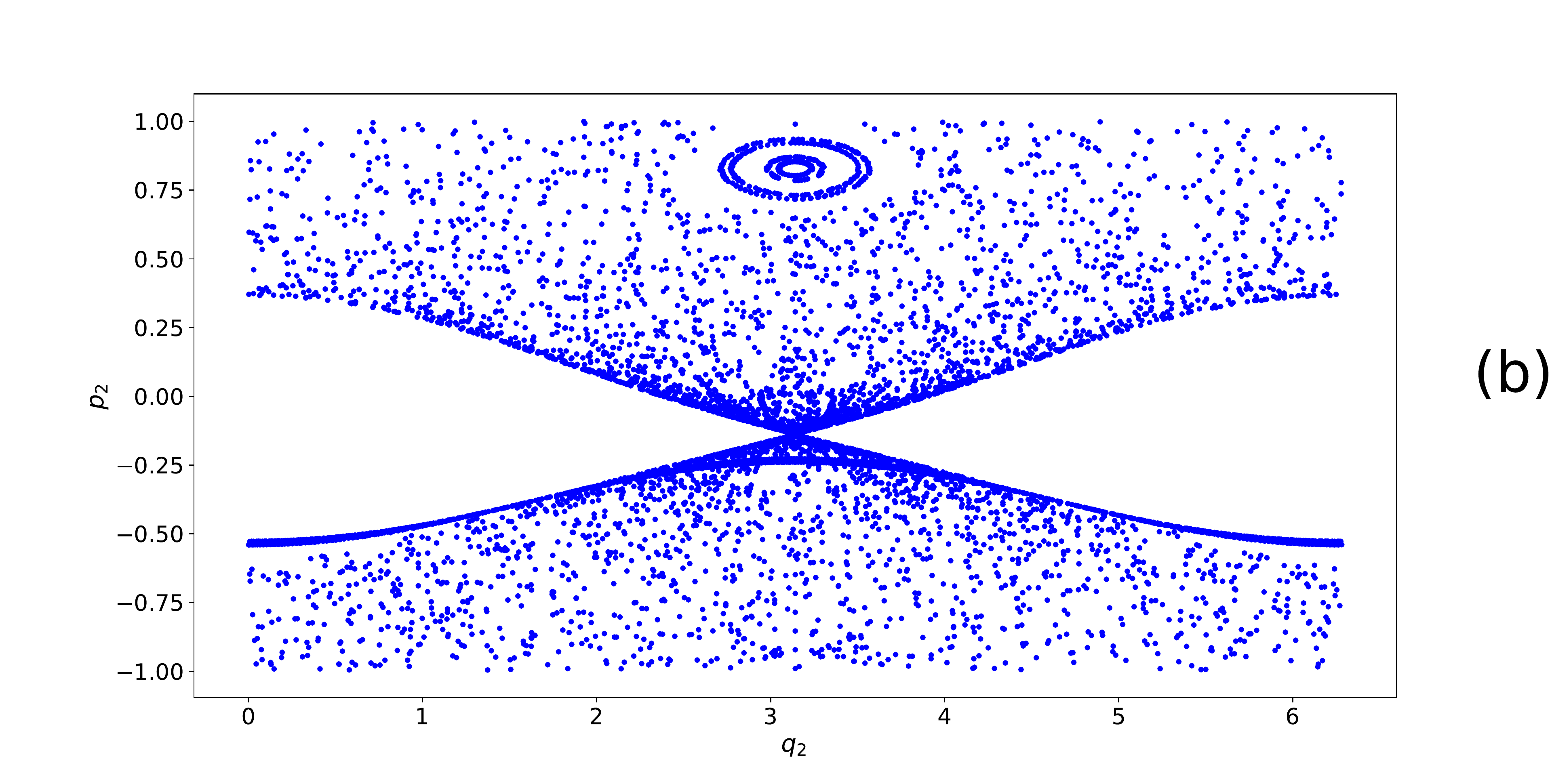}
	\hfill
	\includegraphics[width=0.49\textwidth]{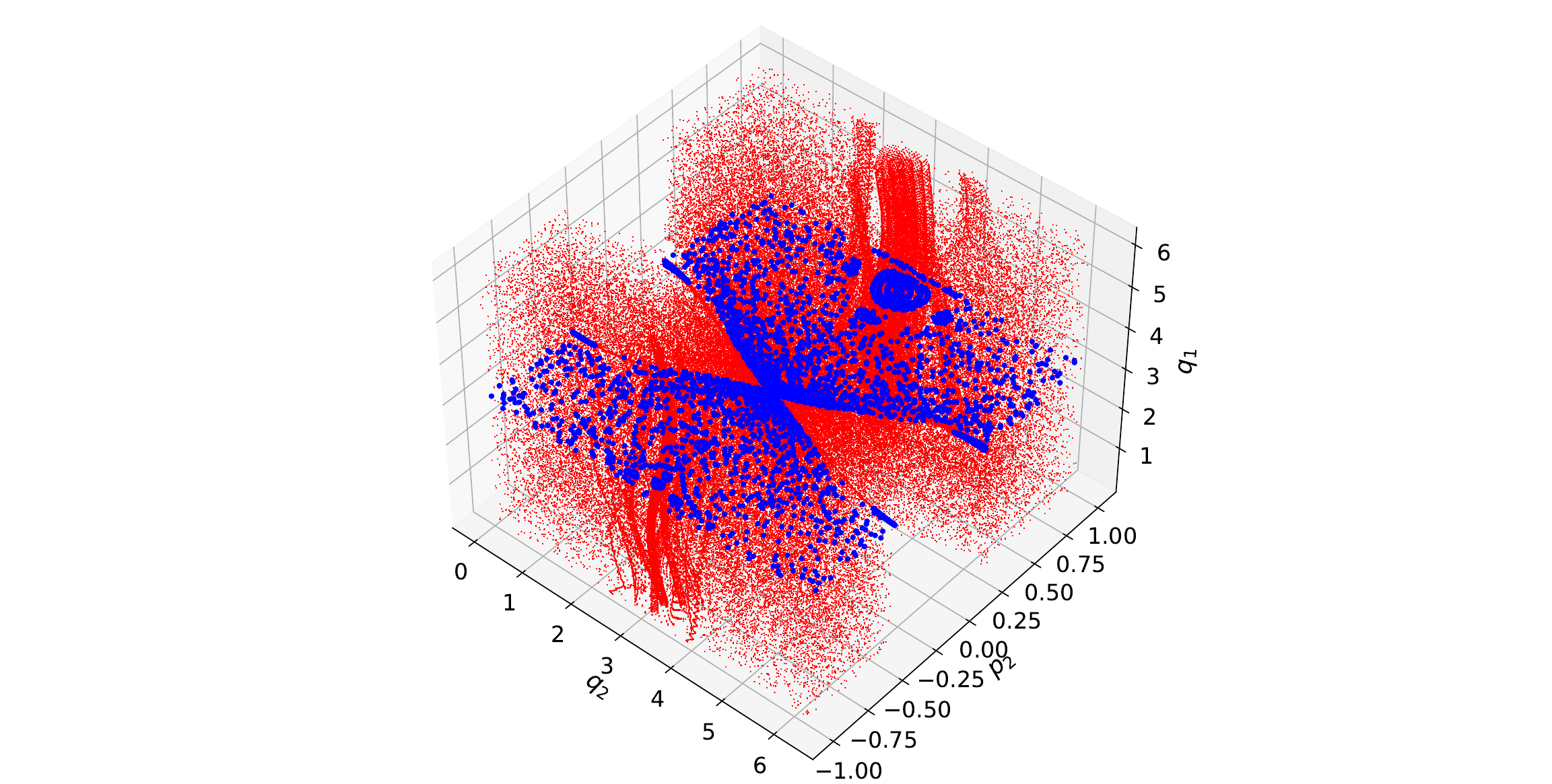}
	\includegraphics[width=0.49\textwidth]{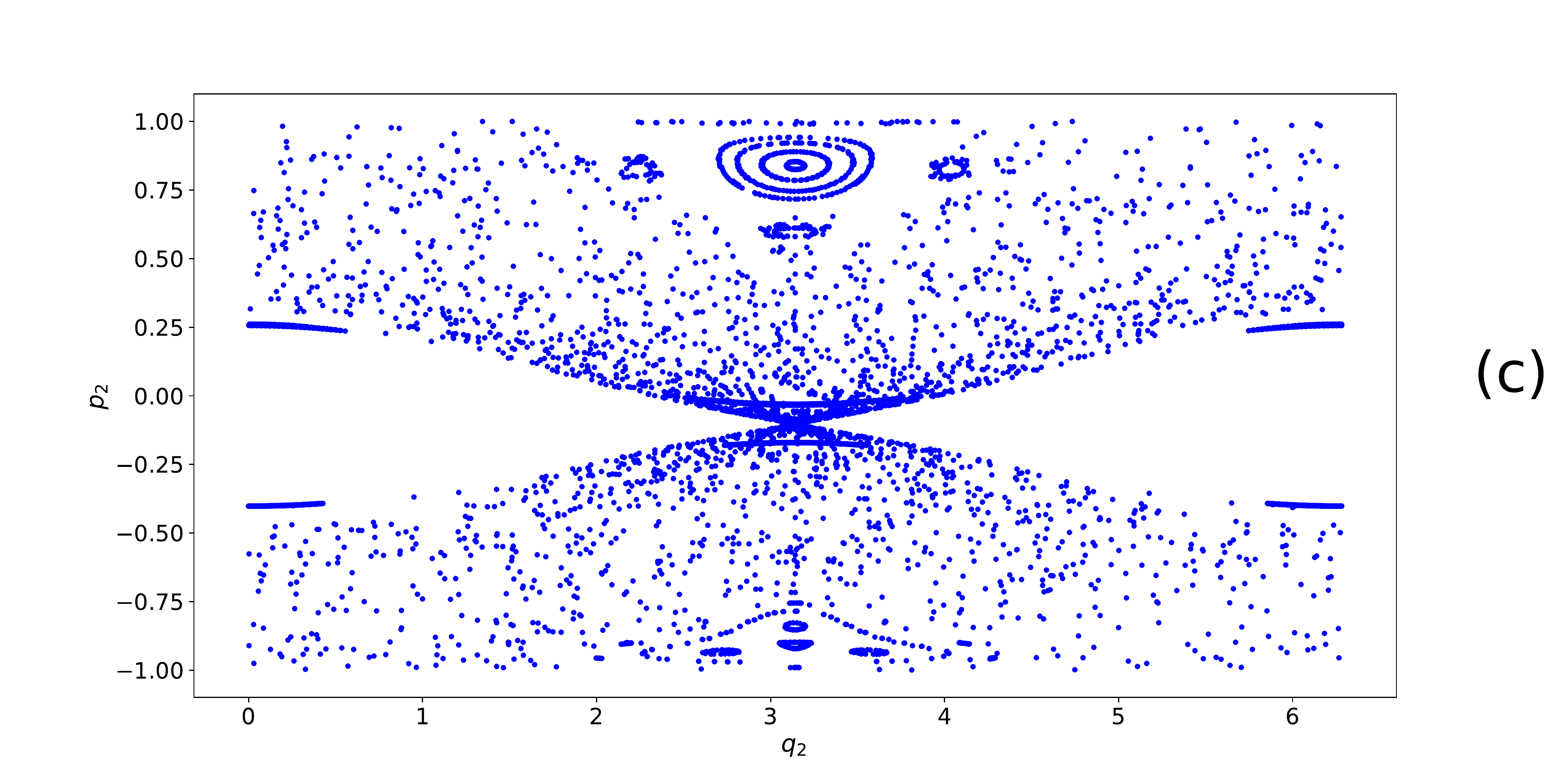}
	\hfill
	\includegraphics[width=0.49\textwidth]{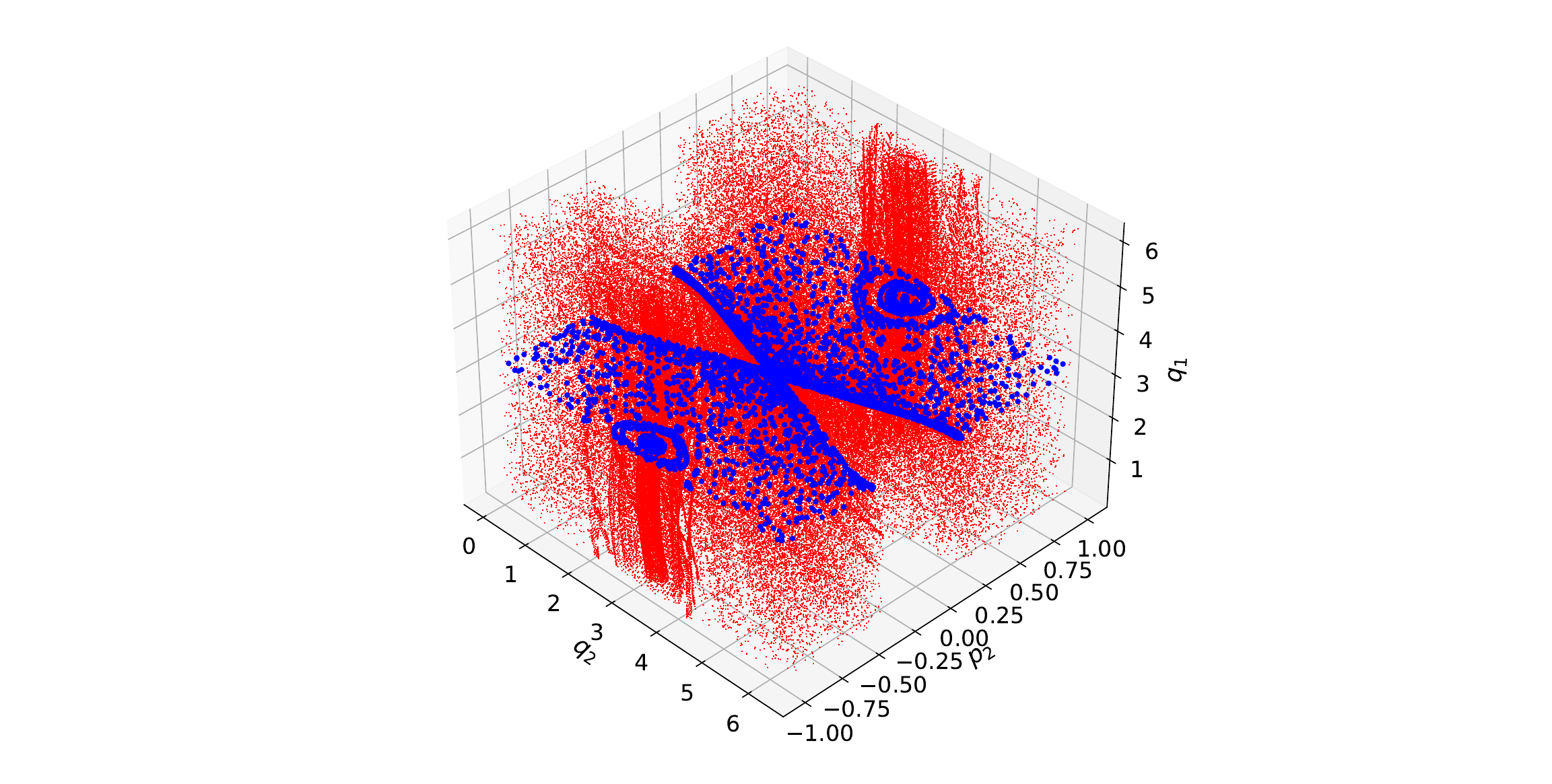}
	\includegraphics[width=0.49\textwidth]{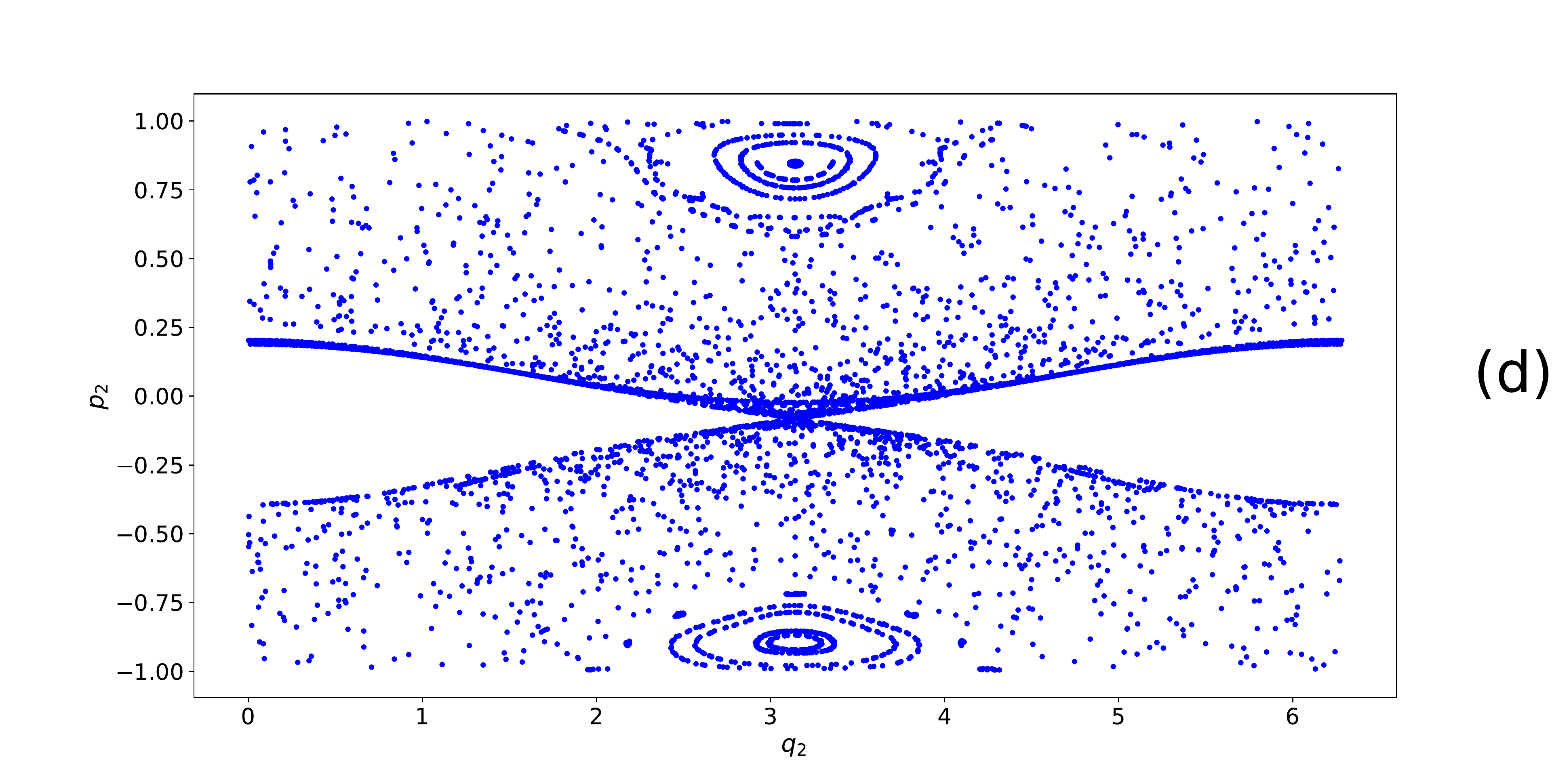}
\caption{Three-dimensional section of energy shells (red) together with Poincaré surfaces of section containing the fixed point in their center. The interaction parameter is monotonically increased from the first to the last plot. The corresponding energies and interactions for each row of plots are given by (a): $E=-0.221$, $J=0.095$, (b): $E=-0.214,~ J=0.156$, (c): $E=-0.21,~J=0.217$ and (d): $E=-0.208,~J=0.278$.}
\label{PhasespacePlots}
\end{figure*}

\begin{figure*}
    \centering
    	\includegraphics[width=\textwidth]{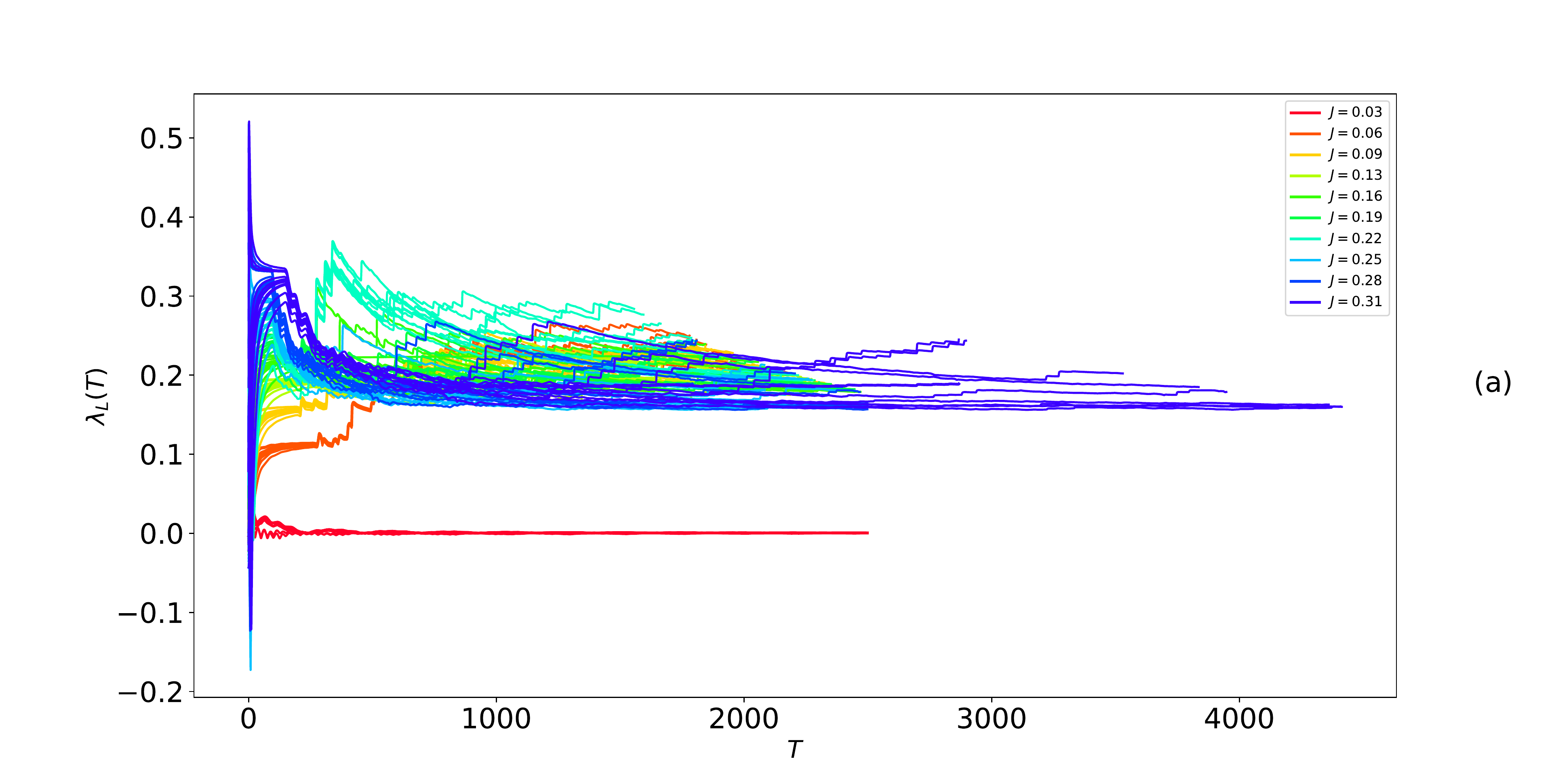}
    	\includegraphics[width=\textwidth]{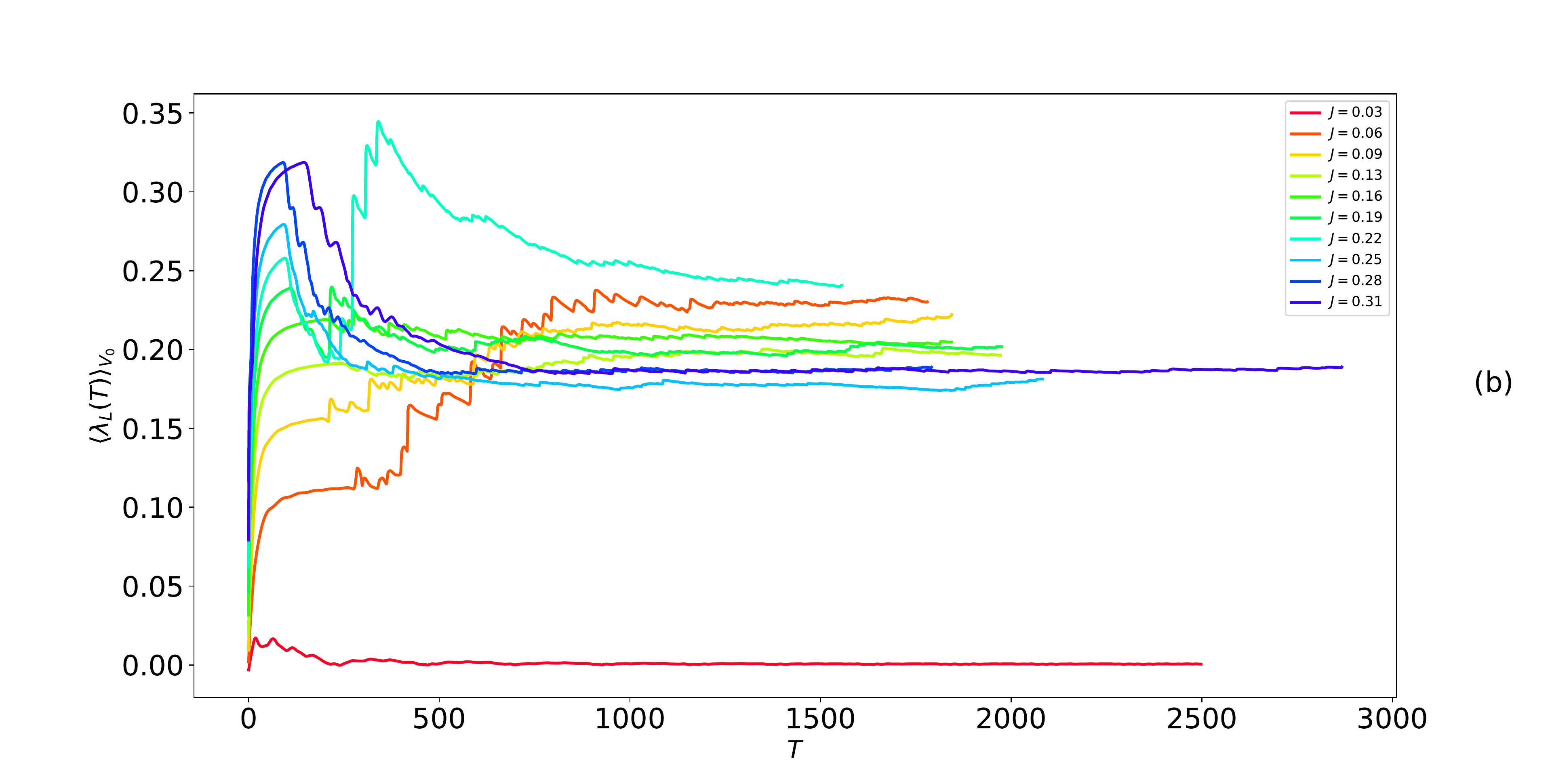}
    \caption{Lyapunov exponent calculation (a) for each of the previously shown phase space regions as well as for intermediate interaction parameters not plotted beforehand. Thin lines of the same color only differ in the used initial deviation vector, thick lines are averages over all thin lines of the same color. For better readability the second plot (b) again shows only the averages used for OTOC comparisons. In both plots $T$ is a large numerical time value up to which the calculation was attempted.}
    \label{LyapunovExponents}
\end{figure*}

In Fig.~\ref{PhasespacePlots} we depict a three-dimensional section of the energy shell together with a two-dimensional Poincaré surface of section for varying interactions.
The computed fixed point is situated in the center of each Poincaré section.
We can clearly see how a chaotic sea appears around the fixed point and for increasing interaction $J$ further regular islands form at the edges.
We see the formation of said islands also in the calculation of the Lyapunov exponents~\cite{Meiss, Froeschle}, since they do not monotonically increase with $J$ but reach a lower positive value once the chaotic part of the region shrinks.
For each interaction parameter we use the method from the previous chapter to compute the Lyapunov exponent of the same deviation $z_0(J)+\epsilon$ from the fixed point for one of ten different deviation vectors $V_0\in\mathrm{T}_{z_0(J)+\epsilon}\mathcal{P}$ and finally average over all those $V_0$ to obtain the final result.
For each interaction the calculations for different deviation vectors are shown in the same color, while the thick lines in the following plot Fig.~\ref{LyapunovExponents} are the averages over all curves of the same color.
This methodology is applied to give an estimate for the dependence on initial deviation vectors as well as to eliminate errors that might be introduced if we accidentally were to choose a deviation vector for which the convergence of the Lyapunov exponent calculation progresses to the correct value more slowly or for which it might even fail.
From the fact that the individual thin curves in Fig. \ref{LyapunovExponents} end at different time values, we see that the divergence of the norm of the curve $\tilde{V}(T)$ can lead to values too large to be represented in the numerics at different times depending on the choice of the initial deviation vector, which further affirms to not rely on a single choice of the former.
The possible numerical error of the calculation is estimated from the energy conservation of the calculation of classical trajectories, which is found to be only of order $10^{-6}$ and is found in App. \ref{appendix2}. 
Other methods for computing Lyapunov exponents such as the "two-particle-method"~\cite{TwoParticleMethod} that does not suffer from the problem of diverging norms was not employed for a number of reasons.
First it is not straightforward to generalize to a non-Cartesian phase space $S^2\times S^2$ since it makes use of both a metric on the space as well as a norm for tangent vectors, which can be naturally obtained from just the Euclidean metric in the case of $\mathcal{P}=\mathbb{R}^{2f}$ but have to be chosen carefully in a more general setting.
Secondly it has been shown \cite{TwoParticleMethodProblems,TwoParticleMethodProblems2} that this method can yield wrong results in certain scenarios including our type of phase space.

With the local and Lyapunov stabilities at hand we are ready to compare with the growth rates of corresponding OTOCs which we compute for the same system parameters and at spin $s=60$ for a two particle system.
Higher spin representations are desirable to represent an as accurate as possible version of a semiclassical limit, but for the two spin system with Hilbert space dimension $\mathrm{dim}(\mathcal{H})=(2s+1)^{2}|_{s=60}=14641$ we already approach the upper limit of numerical feasibility within an acceptable amount of computing time.

\begin{figure*}
\centering
	\includegraphics[width=\textwidth]{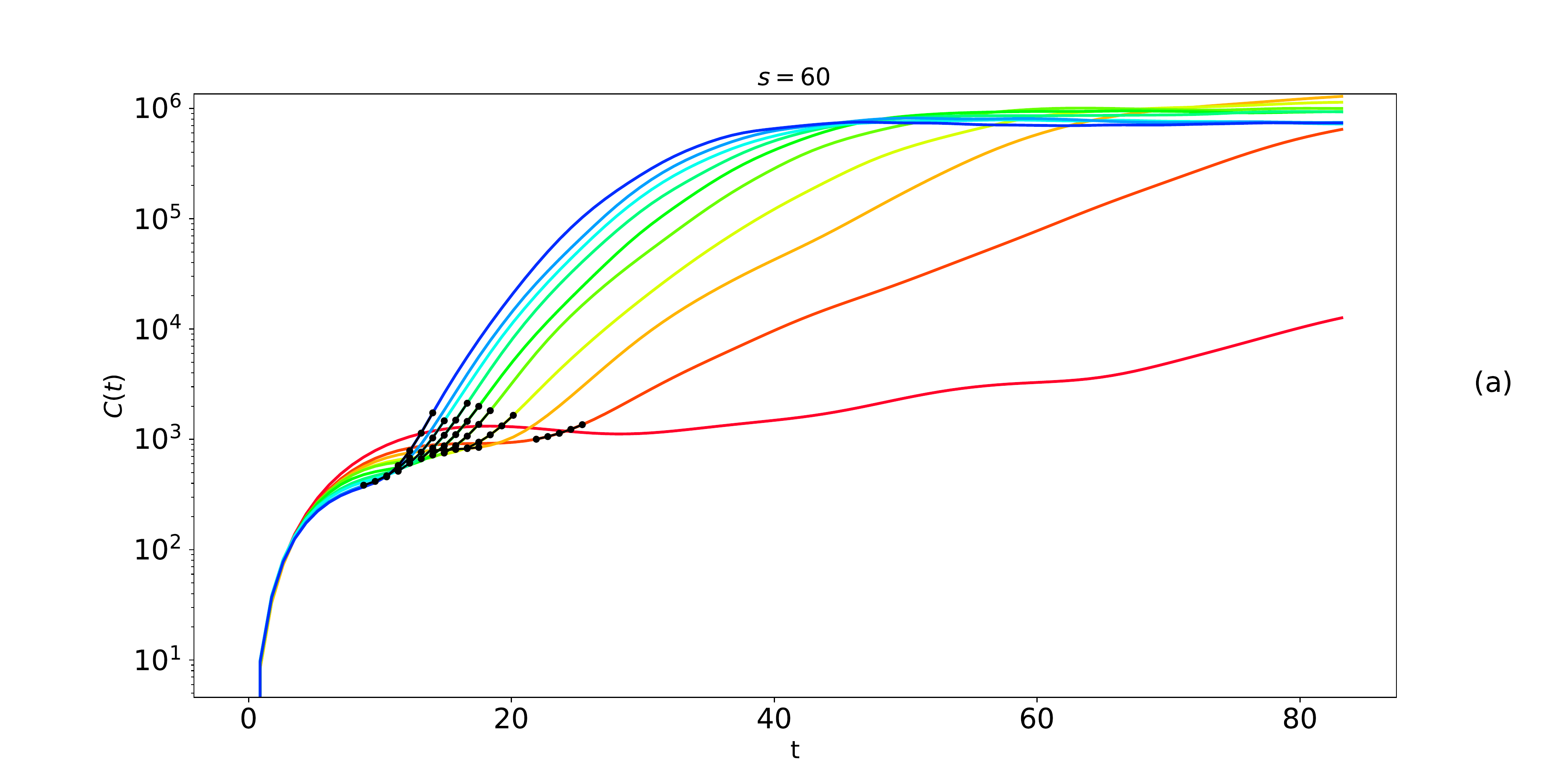}
	\includegraphics[width=\textwidth]{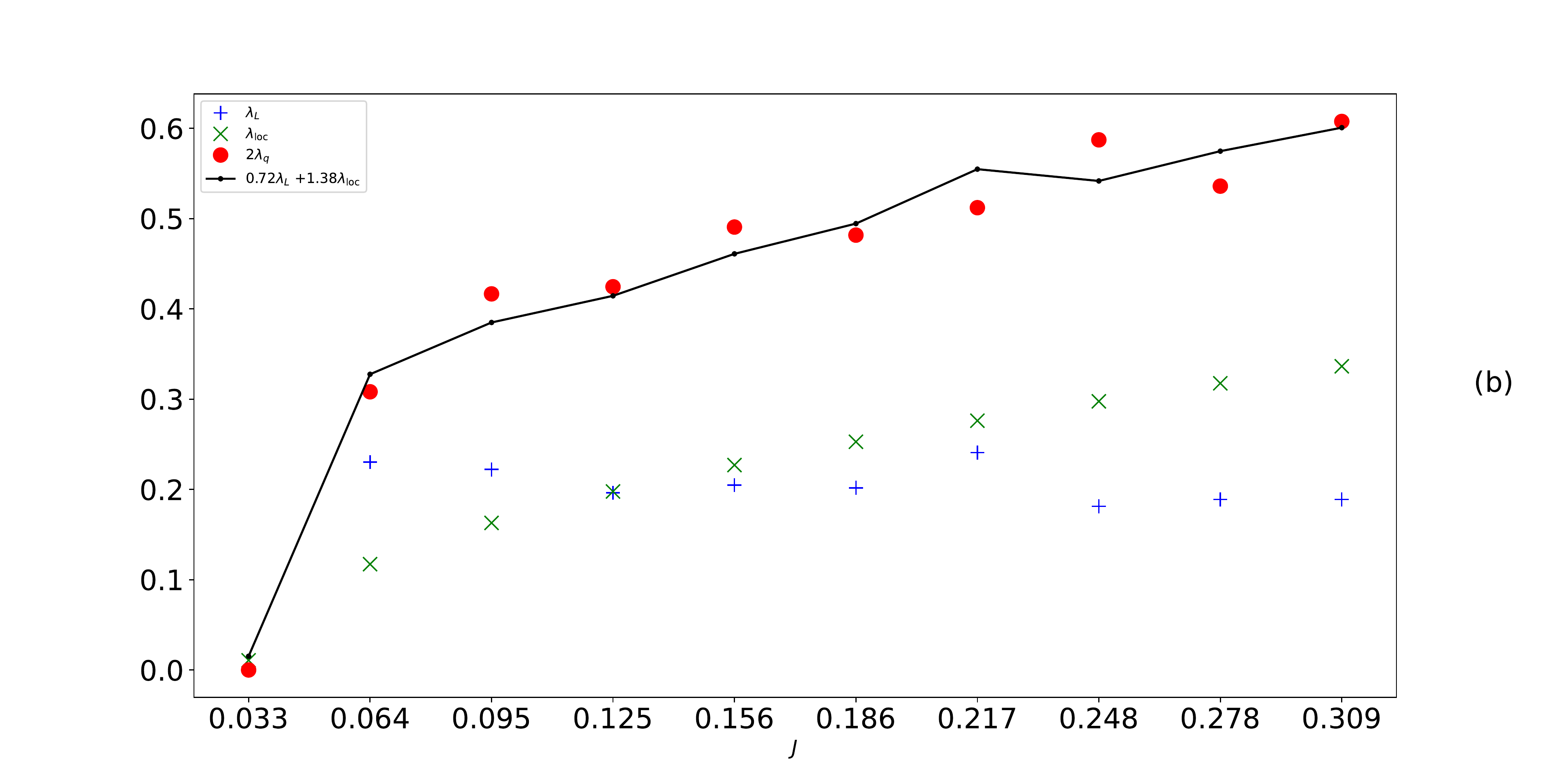}
	\caption{OTOCs for increasing interaction $J$ and coherent state centered at classical fixed point (a) and comparison with classical quantities (b). The OTOC curve for the lowest interaction in (a) corresponds to the integrable region and due to both classical stability quantifiers being very small, this configuration does not display strong enough exponential behavior for performing a fit. The corresponding OTOC exponent is set to zero and not included in the calculation of the parameters $a,b$.}
\label{OTOC_at_fp}
\end{figure*}

For a system with mixed phase space as well as multiple instability classifiers $\lambda_{\text{loc}},\lambda_{\rm L}$ we can not make use of the usual time domain definitions $t_0=1/\lambda_{\rm L}$. 
We instead take a classical state within the chaotic sea and away from the fixed point, compute the coherent state OTOC for it and then look for the temporal region $[t_0,t_1]$ that fulfills the aforementioned theoretical expectation for purely chaotic systems.
For a high enough spin representation the influence of the local instability should not contribute to the OTOC growth here and said expectation is expressed by an exponent fulfilling $\lambda_{\rm q}=\lambda_{\rm L}$ \cite{Maldacena}, which we find for a fixed interaction.
The used fitting function is given by
\begin{align}
\label{Definition_fitfunction}
	g_{\rm q}(t):=a+b\exp(ct)\,\,,
\end{align}
where $c\approx 2\lambda_{\rm q}$.
Afterwards we use this time interval that fulfilled the $\lambda_{\rm q}=\lambda_{\rm L}$ assumption to fit the OTOCs for all further interaction values as seen in Fig.~\ref{OTOC_at_fp} for~$s=60$ together with exponential fits with the aforementioned function $g_{\rm q}(t)$, which is represented as black dots. 
Now the coherent state is centered on the fixed point.
The numerics is performed via Python, using the QuTip package \cite{Qutip,Qutip2}.
In the overview b   elow the OTOC plots we depict the exponent obtained from the fits, the classical parameters as well as the best linear combination of the latter that approximates the OTOC growth rate.
Hence when the black curve overlaps with the red dots representing $2\lambda_{\rm q}$, the assumption~$2\lambda_{\rm q} \approx f_{\rm q}(\lambda_{\rm L},\lambda_{\text{loc}}):= a \lambda_{\rm L} + b \lambda_{\text{loc}}$ for fixed $a,b\in \mathbb{R}$ is fulfilled, where $f_{\rm q}$ is a specific manifestation of the function in the hypothesis~ Eq.~\eqref{hypothesis}.
We fit the data sets  $\{\lambda_{\rm L}(J)\},\{\lambda_{\text{loc}}(J)\}$ with $J$ dependent entries as a whole to the complete data set $\{2\lambda_{\rm q}(J)\}$ to obtain the values for $a,b$ instead of fitting individually for each $J$, which would result in a $J$ dependent $a(J),~b(J)$.
We also tested various non-linear combinations of $\lambda_{\rm L}$ and $\lambda_{\text{loc}}$ but found the simple linear combination depicted in the plots to give the best agreement with the actual OTOC data.

Furthermore, any non-linear combination of classical quantities has the problem that it requires dimensionful fit parameters to result in a dimensionless object as the complete exponent.
However, ignoring this problem of dimensions, we also tested functions like $f_{\rm q}(t)=a \lambda^n_{\rm L} + b\lambda^m_{\rm loc}$ for various integers $n,m\in\mathbb{N}$ which all turned out to yield worse results than $n=1,m=1$.
Additionally, only a linear combination directly reduces to known extreme cases for either $a=0$ or $b=0$.

\begin{figure*}
\centering
	\includegraphics[width=\textwidth]{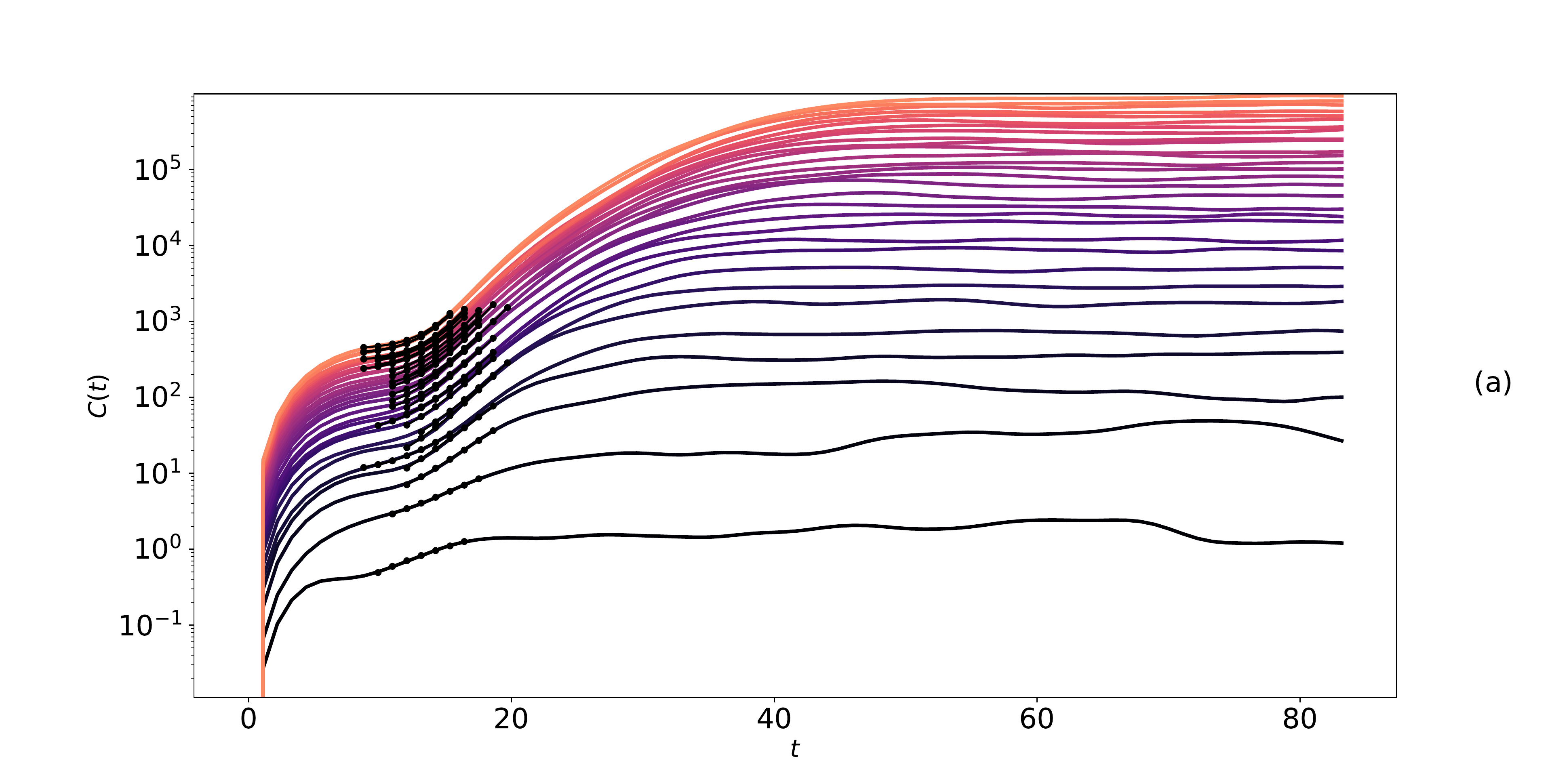}
	\includegraphics[width=\textwidth]{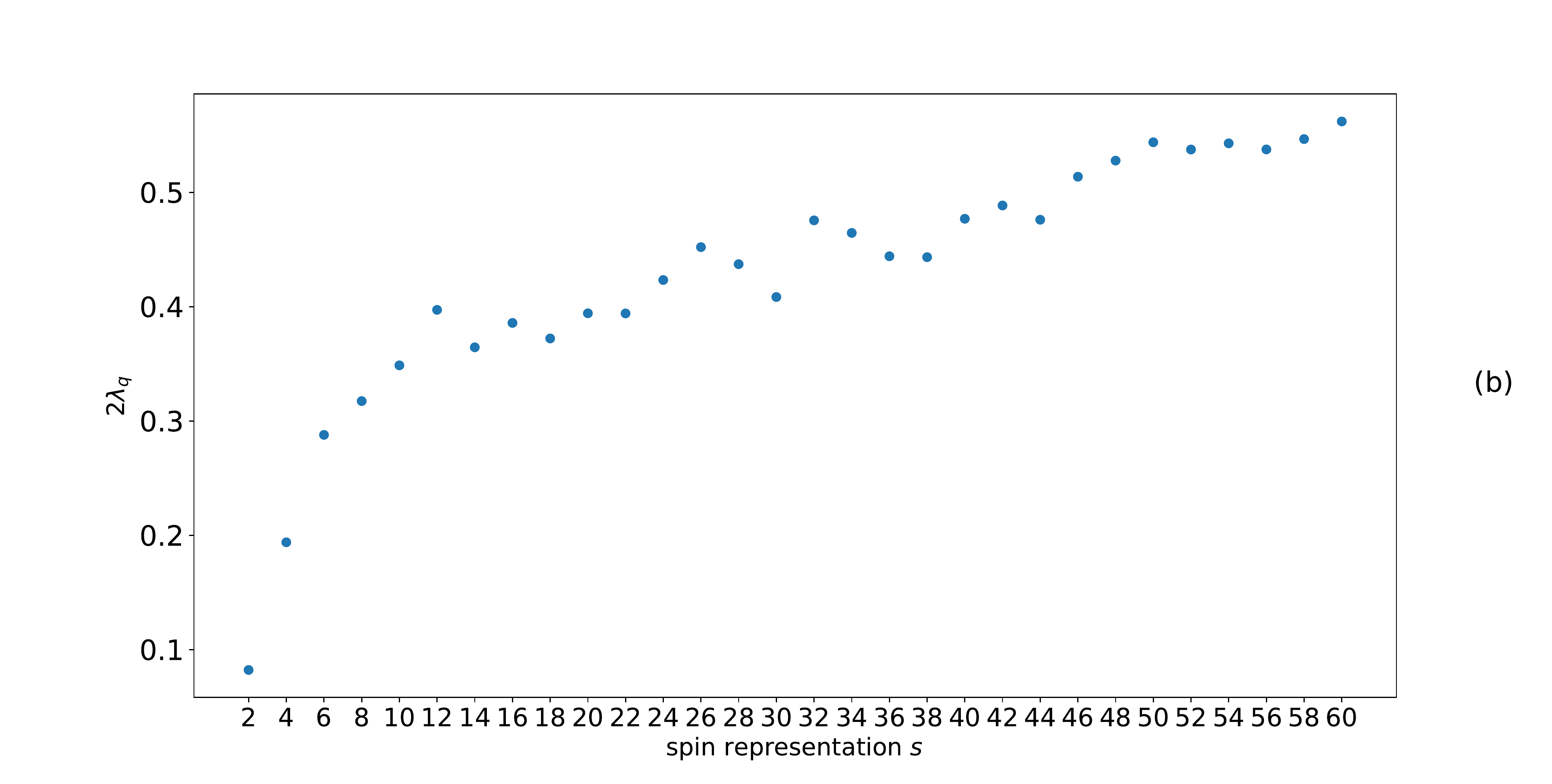}
	\caption{OTOCs (a) and their growth rates (b) depending on spin $s$, fixed interaction $J=0.217$ and coherent state centered at classical fixed point. Each value of (b) corresponds to one exponential to a curve in (a), performed in the same way as previously described for the $J$-dependent fits.}
\label{OTOC_spin_dependence}
\end{figure*}

\begin{figure*}
\centering
	\includegraphics[width=\textwidth]{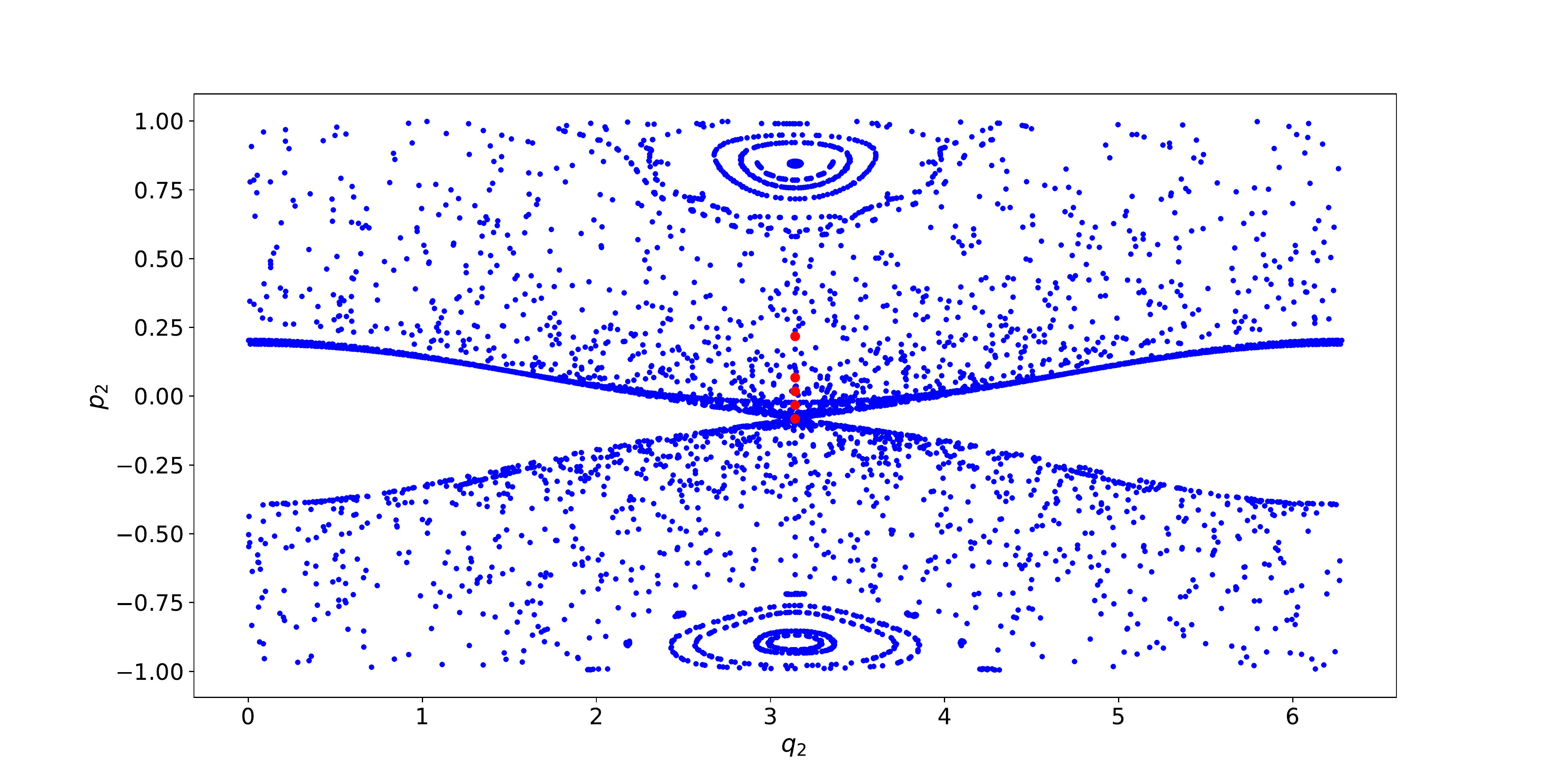}
	\caption{The red points lie on the previously defined line and are used as centers of the coherent states for OTOC calculations. }
\label{Line_in_Phasespace}
\end{figure*}

\begin{figure*}
\centering
	\includegraphics[width=\textwidth]{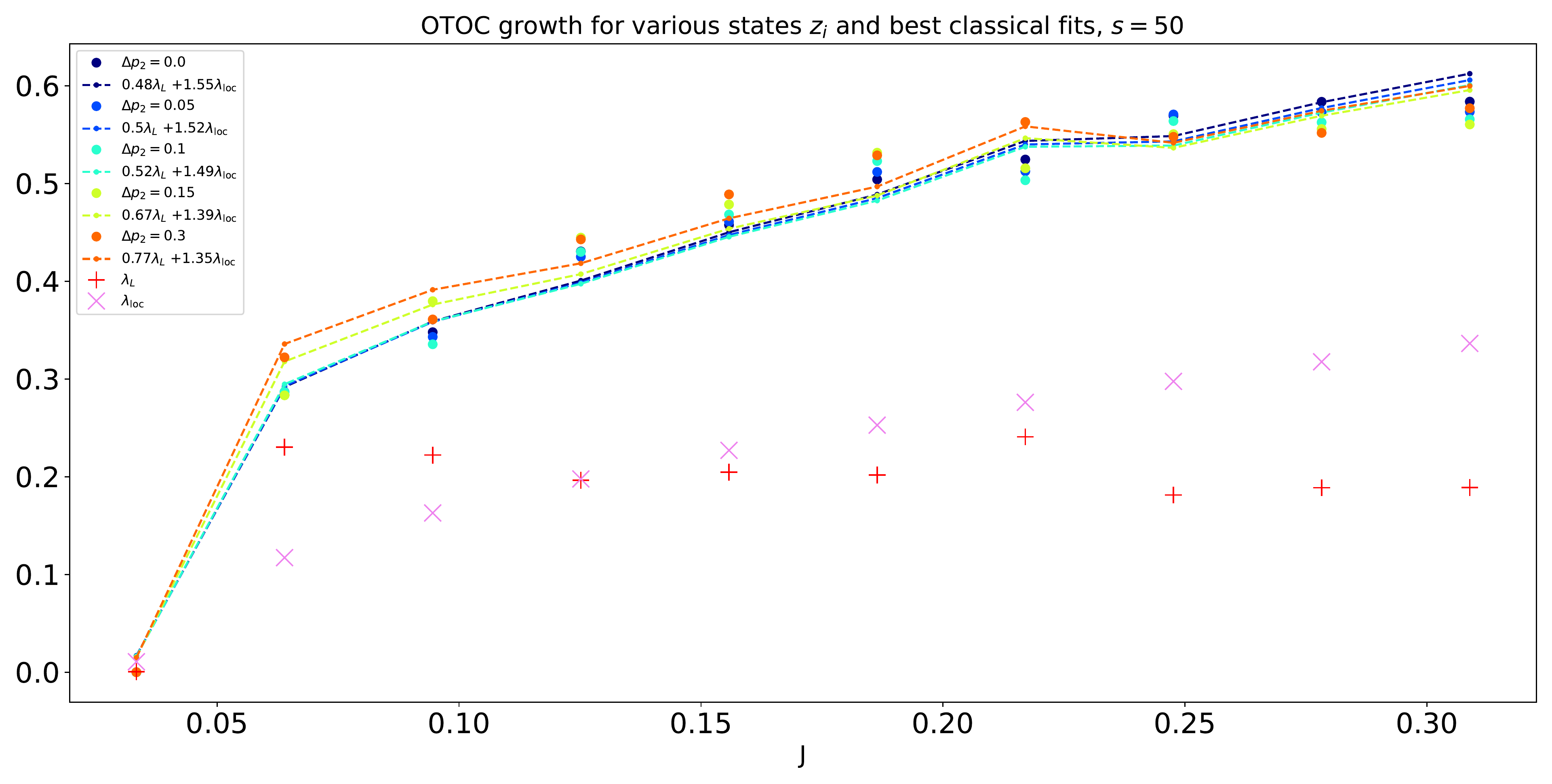}
\caption{Results for the OTOC fits at various points $z_i$ along the line in phase space Fig. \ref{Line_in_Phasespace}. The large dots represent $2\lambda_{\rm q}$.}
\label{results_overview_final}
\end{figure*}

We see that the Lyapunov exponent alone can not explain the OTOC growth as in the case that was used to fix the time interval.
By fixing the interaction again and recalculating the OTOC for this single interaction value now in dependence of the spin representation we can assess whether the exponential growth rate is already described by semiclassics, which is the case once its variance becomes small.
The corresponding spin dependence of the OTOCs as well as their exponential fits are shown in Fig.~\ref{OTOC_spin_dependence}.
Unlike in~\cite{LyapunovQMSpinChains} our definition of the OTOC theoretically results in a spin-independent exponential growth rate instead of only reaching one asymptotically.
Deviations from this expectation are not caused by a spin dependence $\lambda_{\rm q}(s)$, but by the fact that reaching the semiclassical domain itself depends on the spin representation.
From this spin dependent determination of the OTOC growth we see that from around $s=50$ fluctuations become relatively small, allowing us to reduce the numerical effort to this smaller value for the following calculations.
In order to test the dependence of the OTOCs on the classical quantities in more depth, we pick further classical states along a line in phase space which originates at the fixed point and reaches into the purely chaotic region above it as depicted in Fig. \ref{Line_in_Phasespace}.
For each of the points on the line above procedure is repeated for spin $s=50$ rendering possibly a comparison of the fit parameters $a,b$ in dependence of the classical state used for the computation.
For the fit of $a$ and $b$ we still use the same local and global stability quantifiers $\lambda_{\text{loc}},~ \lambda_{\rm L}$ as before.
One might expect that the coherent state, due to being only slightly spread out already at time $t=0$ with a radius $1/s|_{s=50}=0.02$ in phase space does no longer include the fixed point already for the first deviation of $\Delta p_2=0.05$.
Then it would only include the fixed point again as the chaotic dynamics makes it spread out over the chaotic sea.
The temporal region is now no longer determined statically by comparing with the $\lambda_{\rm q}\approx \lambda_{\rm L}$ condition, but instead we optimize it for the best possible exponential fit, by splitting the time axis into small intervals, fitting for each of them and filtering out the optimal fit from the results.
This procedure also explains the disagreement of the fit parameters $a,b$ for the consecutive evaluations from the ones found in Fig. \ref{OTOC_at_fp} for the coherent state centered on the fixed point.
As a consequence, the OTOC should depend less on $\lambda_{\text{loc}}$ and for a point in the chaotic sea and away from the fixed point it should show the growth behavior with just $2\lambda_{\rm L}$.
Hence if we choose the line in phase space as
\begin{align}
\label{Parametrised_path}
	\gamma:&[t_0,t_1] \rightarrow \mathcal{P}\\
		& \gamma(t_0):= z_0,\,\,\, \gamma(t_1):=z_1 \nonumber
\end{align}
we expect
\begin{align}
\label{Expected_fit_behavior}
	& a(\gamma(t))=:a(t),\,\,\, b(\gamma(t))=:b(t)\\
	& a(t_1)=2,\,\,\, b(t_1)=0\,\,\,\text{.}\nonumber
\end{align}
Choosing the point $z_1$ to be just a change along the $p_2$ direction, we can also express this curve in coordinates as
\begin{align}
\label{Parametrised_path_coordinates}
    & z(\gamma(t))=(q_1(t),q_2(t),p_1(t),p_2(t))=\\
    & =\left(q_1(t_0),q_2(t_0),p_1(t_0),p_2(t_0)+\frac{0.3(t-t_0)}{t_1-t_0}\right)\,\,\,,\nonumber
\end{align}
\\
\\
where $z_0\cong(q_1(t_0),q_2(t_0),p_1(t_0),p_2(t_0))=({q_0}_1,{q_0}_2,{p_0}_1,{p_0}_2)$ are the coordinates of the fixed point $z_0$.
In the plot in Fig.~\ref{results_overview_final} we present the results in the same manner as before.

\begin{table}
    \caption{Results from fits. The contributions $a,~b$ are used to approximate the OTOC fits as $a\lambda_{\rm L} + b\lambda_{\text{loc}}$.}
    \centering
    \newcolumntype{P}[1]{>{\centering\arraybackslash}p{#1}}
    \begin{tabular}{P{25mm}|P{20mm}|P{20mm}|P{15mm}}
        deviation $\Delta p_2$ 
        of fixed point coordinate        
     & $\lambda_{\rm L}$ contribution $a$
     & $\lambda_{\text{loc}}$ contribution $~~b$
     & \vfill $a+b$\\
    \hline
    0.00 & 0.48 & 1.55 & 2.03\\
    0.05 & 0.50 & 1.52 & 2.02\\
    0.10 & 0.52 & 1.49 & 2.01\\
    0.15 & 0.67 & 1.39 & 2.06\\
    0.30 & 0.77 & 1.35 & 2.12\\
    \end{tabular} 
\label{Table_results}
\end{table}
By looking at the results in Tab. \ref{Table_results} we can see that our assumption is fulfilled qualitatively in that the dependence of the OTOC on the Lyapunov exponent grows away from the fixed point while the dependence on the fixed point stability shrinks.
However, the best parameters $a,b$ are differing from the expected limit cases $a=2,b=0$ farthest away from the fixed point as well as $a=0,b=2$ right on top of the fixed point by a large margin, which means in particular that the combination of the two classical quantities is better suited to describe the exponential growth of the OTOC than either of them alone.
One curious observation we can make is that the sum $a+b$ of the parameters stays close to 2 for all considered points in phase space.
This might hint towards a conservation or limitation of the total instability experienced by the OTOC.
This numerical observation also points to the idea that in the present scenario (where the finite extension of the phase space representation for any localized state leads to a non-uniform scrambling rate) the natural interpretation of the pre-Ehrenfest quantum Lypaunov exponent is simply a weighted average of the different contributions. Indeed, by writing
\begin{equation}
  a=\frac{a+b}{2}+\frac{a-b}{2}, {\rm \ \ and \ } b=\frac{a+b}{2}-\frac{a-b}{2} 
\end{equation}
and using~$a + b\,\approx \,2$
we get
\begin{equation}
  2\lambda_{{\rm q}}=2\left(\frac{\lambda_{{\rm loc}} +\lambda_{{\rm L}}}{2}\right)+{\cal O}(\lambda_{{\rm loc}} -\lambda_{{\rm L}}),
\end{equation}
while factor the 2 in front simply reflects the second power of the commutator in the definition of $C^{\hat{A}\hat{B}}_{\rho}(t)$. From these considerations, we expect for this sum $a + b$$\,\approx\,$
$n$ if more general correlations functions involving an $n$-th power of operators are used, which might be an interesting investigation for future works. The parameters $a,~b$ themselves are then in general interpreted as classifiers of the interplay between local and delocalized effects on the scrambling.

\FloatBarrier
\section{Testing the Hypothesis for Bose-Hubbard Rings}
\label{ch3_2}
\begin{figure}[h!]
    \centering
    \subfigimg[width=\linewidth]{\hspace{0.825\linewidth}\textbf{a)}}
    {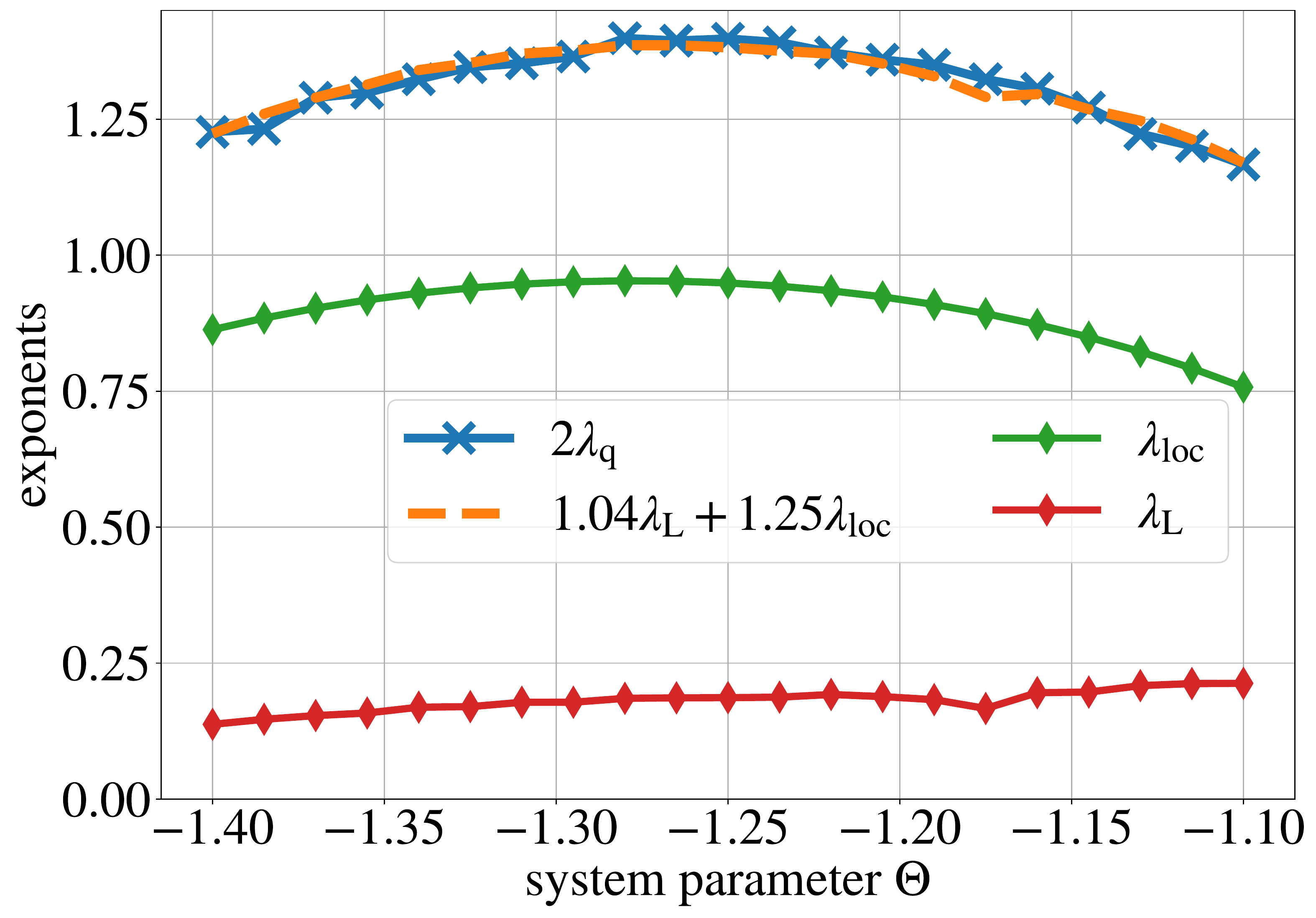}
    \\
    \subfigimg[width=\linewidth]{\hspace{0.825\linewidth}\textbf{b)}}
    {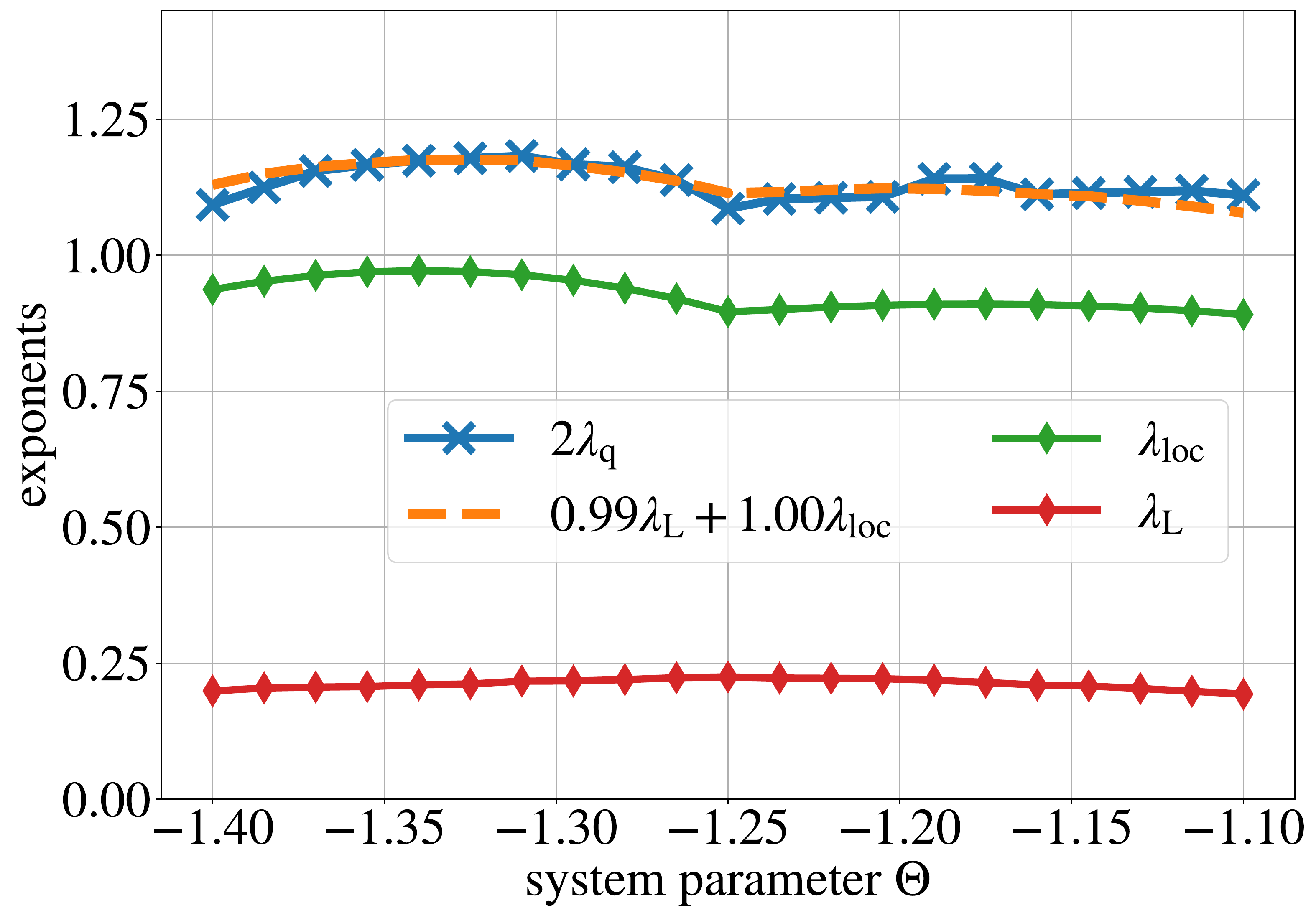}
    \caption{Examination of hypothesis~Eq.~\eqref{hypothesis} for OTOC calculated at the fixed point $z^{\rm H}$: top panel \textbf{a)} 3-site ring; bottom panel \textbf{b)} 4-site ring; in both rings we obtain an excellent agreement for the hypothesis Eq.~\eqref{hypothesis}.}
    \label{fig:BH_FP_exponents}
\end{figure}

\begin{figure}[h!]
    \centering
    \setbox1=\hbox{\includegraphics[width=\linewidth]{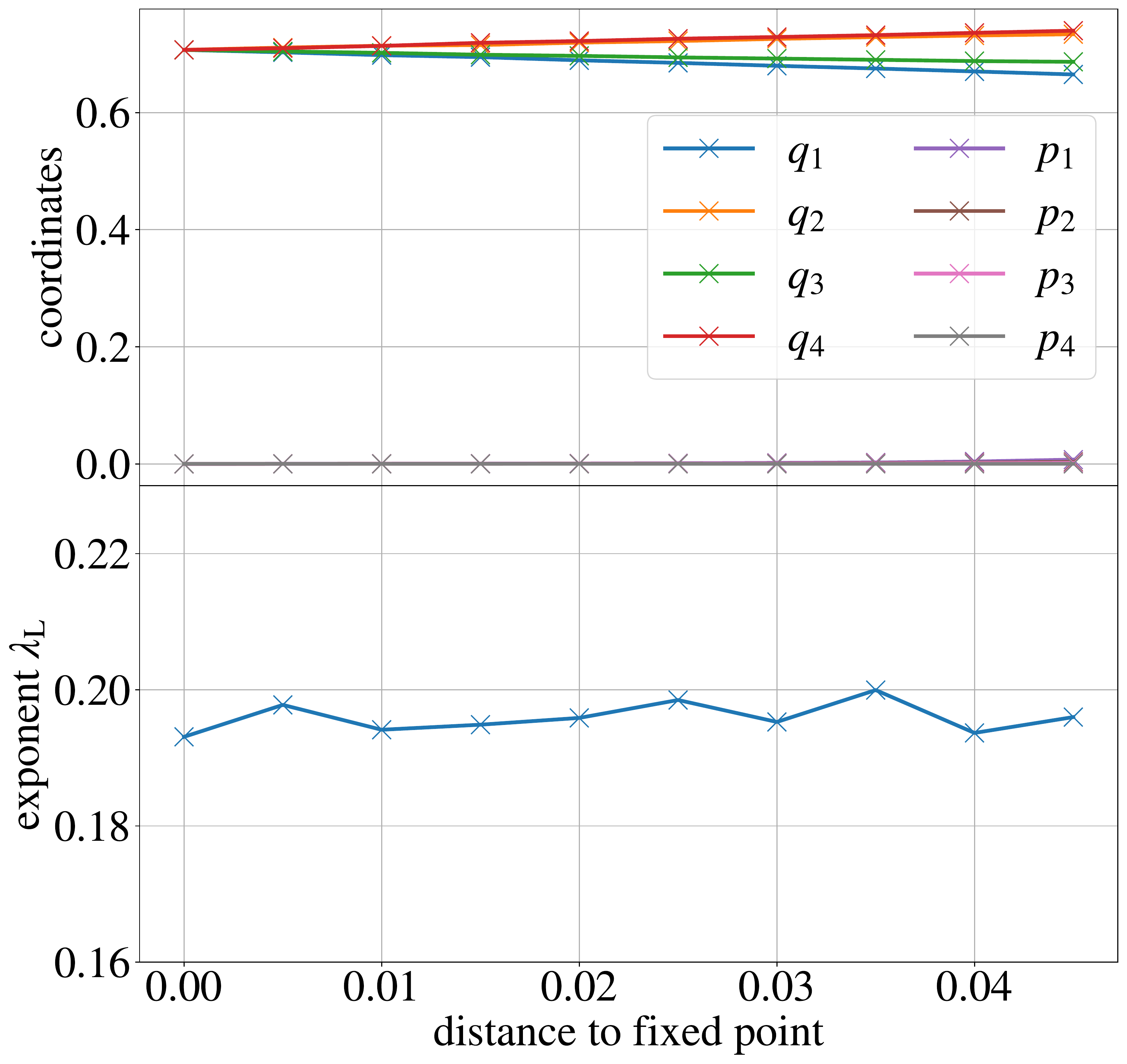}}
    \leavevmode\rlap{\usebox1}
    \rlap{\hspace*{10pt}\raisebox{\dimexpr\ht1-2.5\baselineskip}{\hspace{7.5cm}\textbf{a)}}}
    \rlap{\hspace*{10pt}\raisebox{\dimexpr\ht1-11.2\baselineskip}{\hspace{7.5cm}\textbf{b)}}}
    \rlap{\hspace*{10pt}\raisebox{\dimexpr\ht1-18.85\baselineskip}{\hspace{7.55cm}\phantom{11}}}
    \phantom{\usebox1}
    \caption{Moving away from the homogeneous fixed point for $L=4,~N=40$ and $\Theta=-1.1$; top panel \textbf{a)}: we fix the energy and decrease the $q_{1}$ coordinate from $\sqrt{2/L}$; the coordinates $p_{1},p_{2},p_{3},p_{4}$ lay on top of each other; bottom panel \textbf{b)}: the Lyapunov exponent remains constant indicating we remain inside the chaotic region around the homogeneous fixed point~$z^{\text{H}}$.}
    \label{fig:BH_coordinates}
\end{figure}

\begin{figure*}[h!]
    \subfigimg[width=0.5\linewidth]{\hspace{1.1cm}\textbf{a)}}{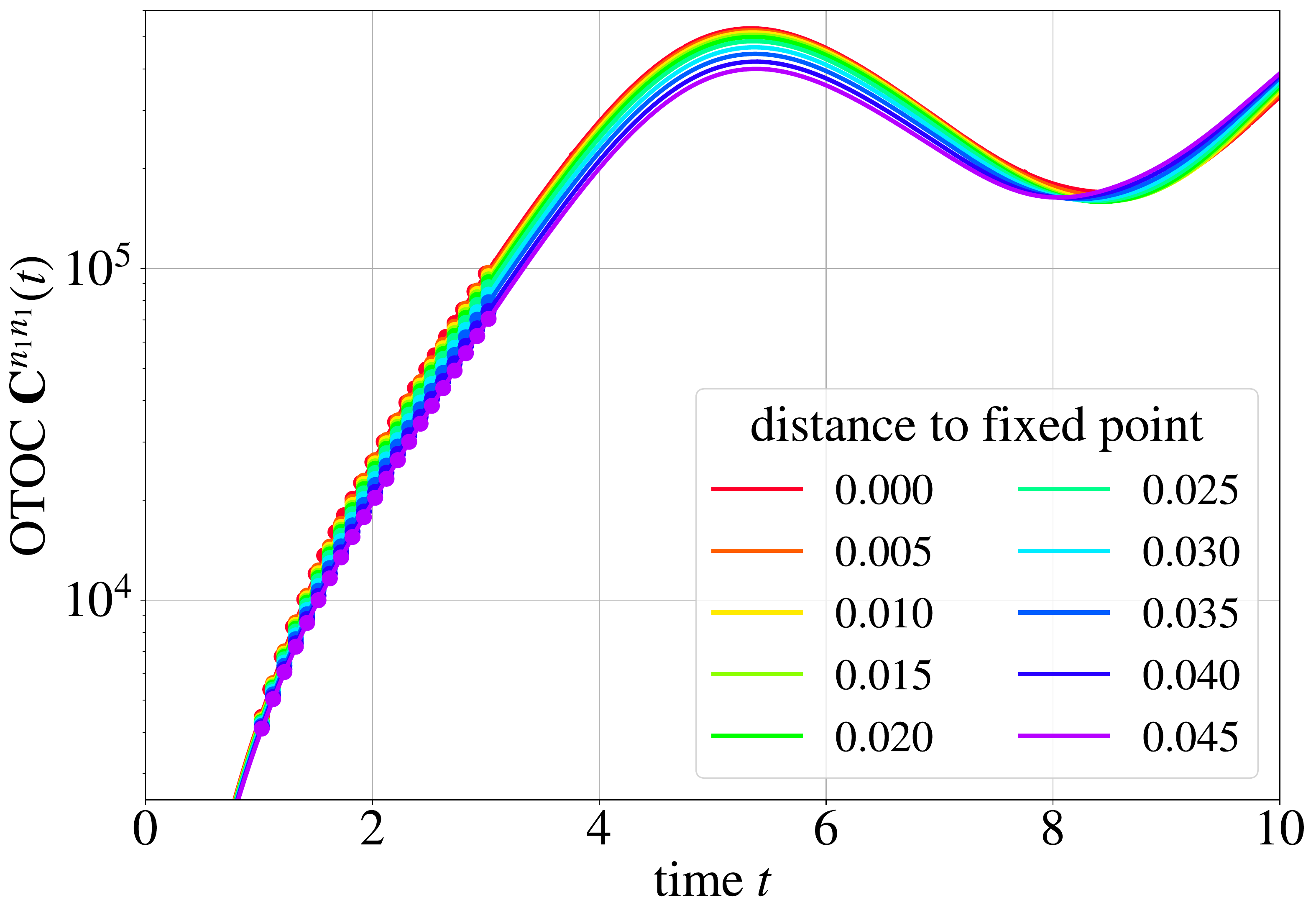}
    \subfigimg[width=0.5\linewidth]{\hspace{1.1cm}\textbf{b)}}{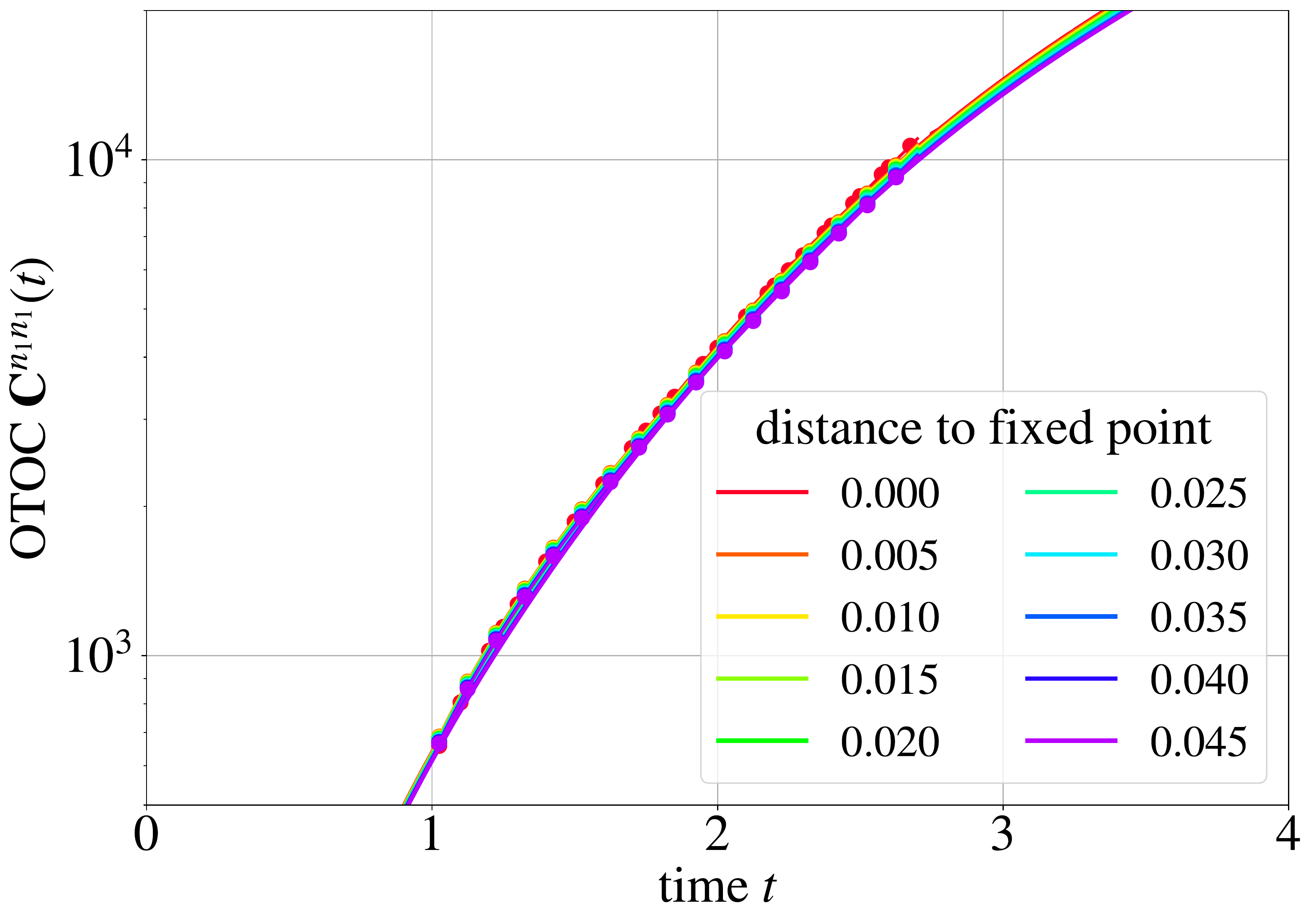}%
    \\
    \subfigimg[width=0.5\linewidth]{\hspace{1.1cm}\textbf{c)}}{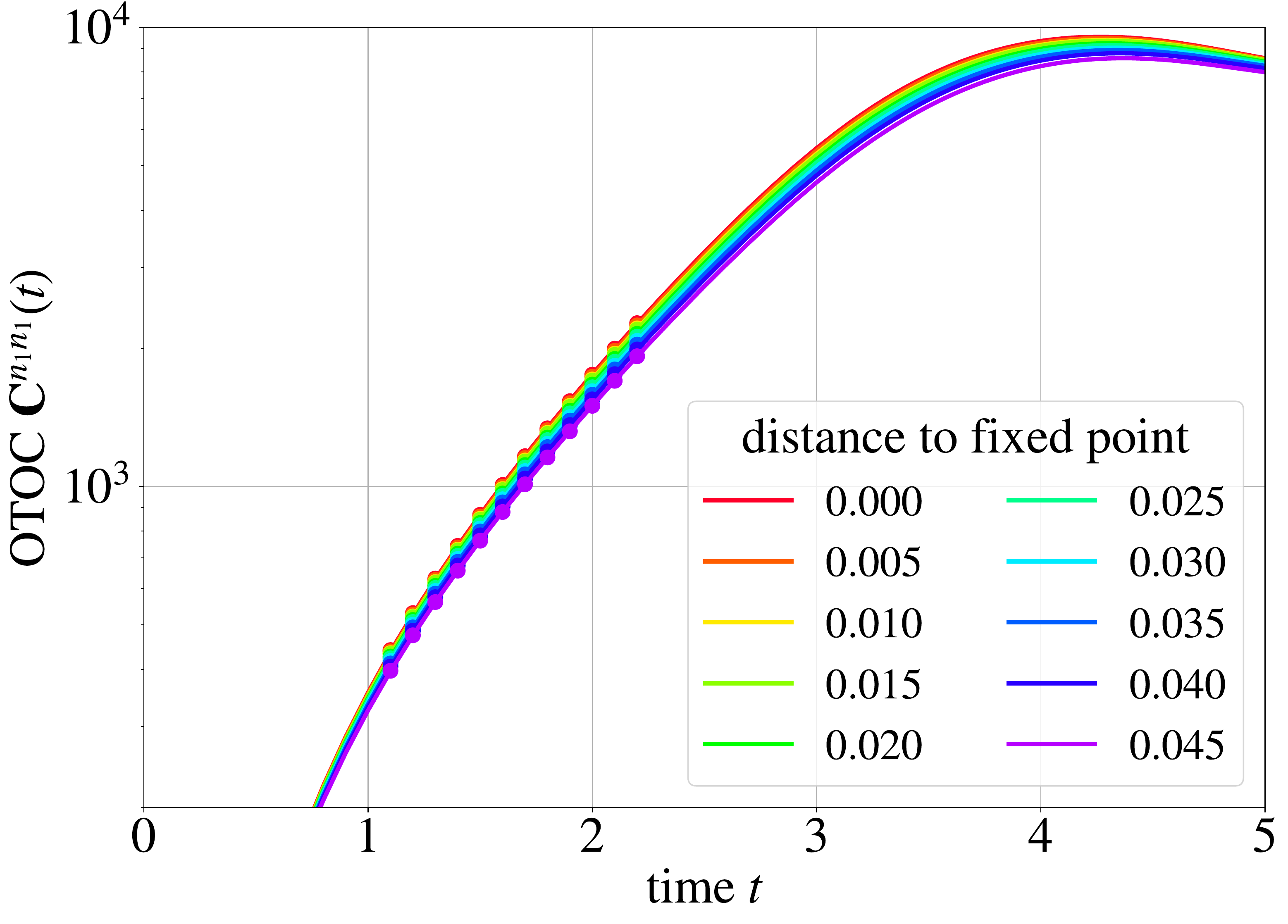}
    \subfigimg[width=0.5\linewidth]{\hspace{1.1cm}\textbf{d)}}{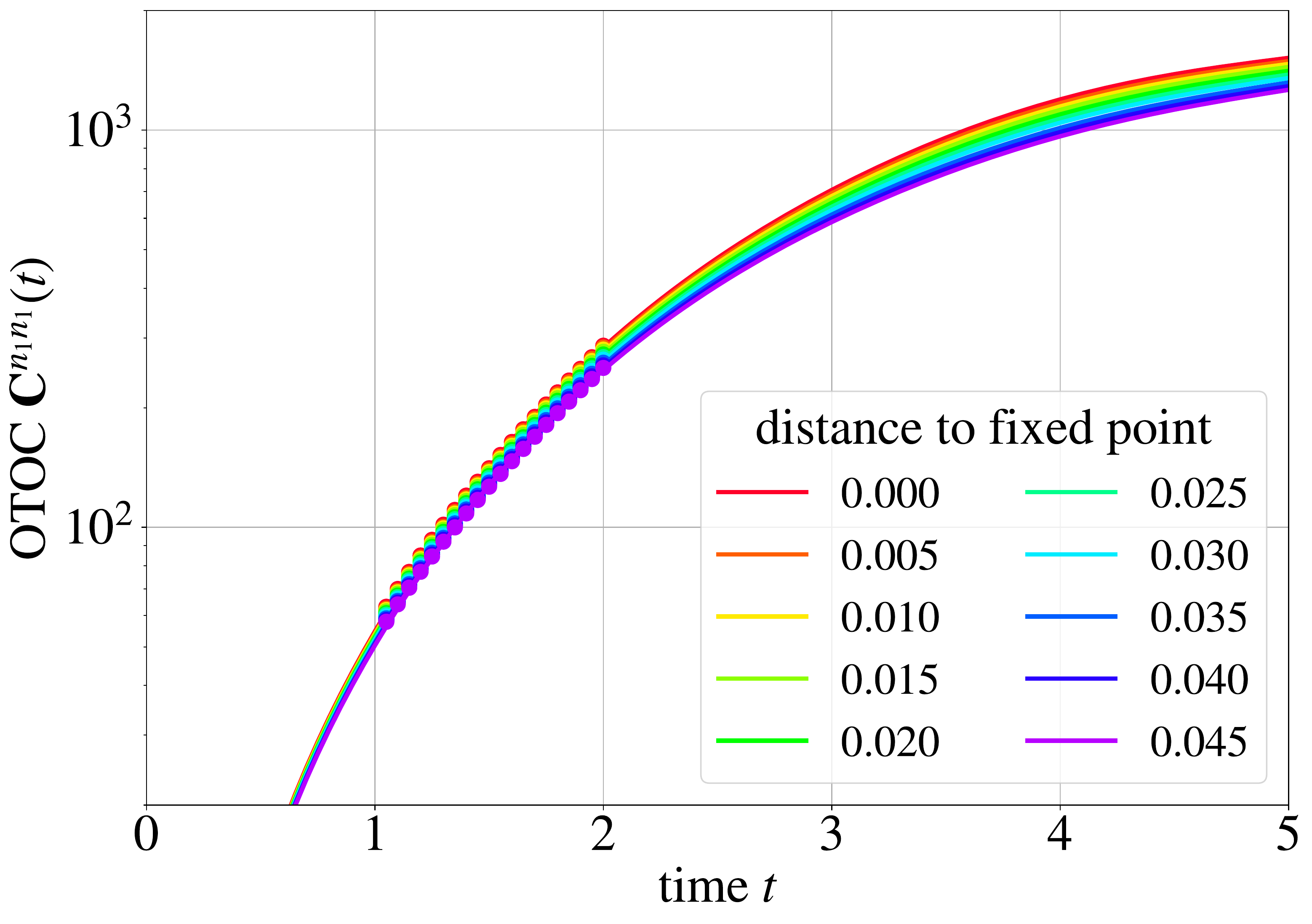}%
    \caption{OTOC plots for $L=3$, $N=100$ with $\Theta=-1.1$ (panel \textbf{a)}), for $\Theta=-1.1$ (panel \textbf{b)}) and for $L=4$, $N=40$ with $\Theta=-1.1$ (panel \textbf{c)}), for $\Theta=-1.1$ (panel \textbf{d)}); color encodes the distance to the fixed point $z^{\rm H}$, dotted points mark the fitted exponential function $g_{\rm q}(t)$~Eq.~\ref{Definition_fitfunction}. OTOCs tend to decrease with the distance to the fixed point or to be highly stagnant (e.g., panel \textbf{b)}.  
    }
    \label{fig:BH_plot_OTOCs}
\end{figure*}

\begin{figure*}[h!]
    \centering
    \subfigimg[width=0.75\linewidth]{\hspace{1.5cm}\textbf{a)}}{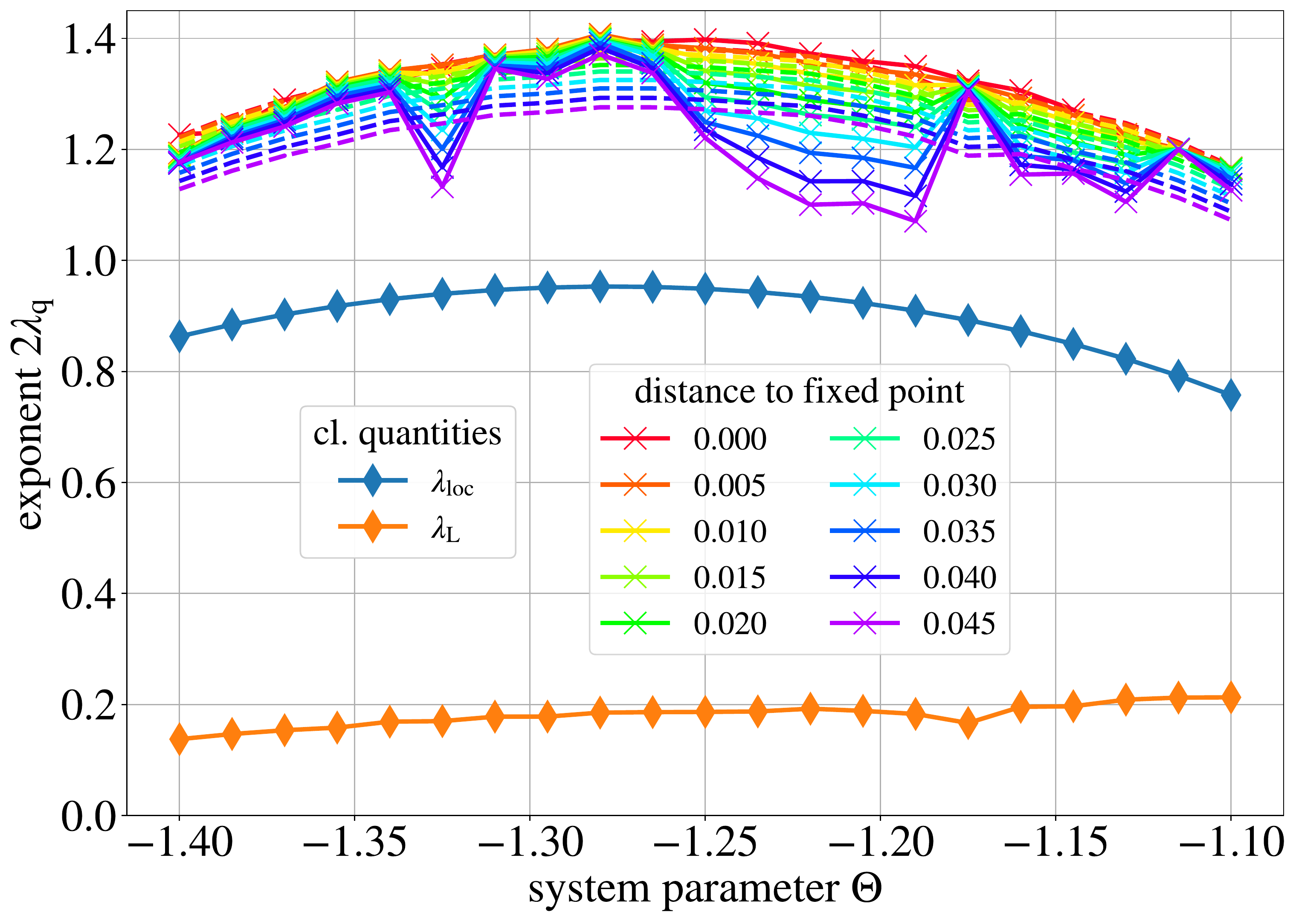}%
    \\
    \subfigimg[width=0.75\linewidth]{\hspace{1.5cm}\vspace{0.5cm}\textbf{b)}}{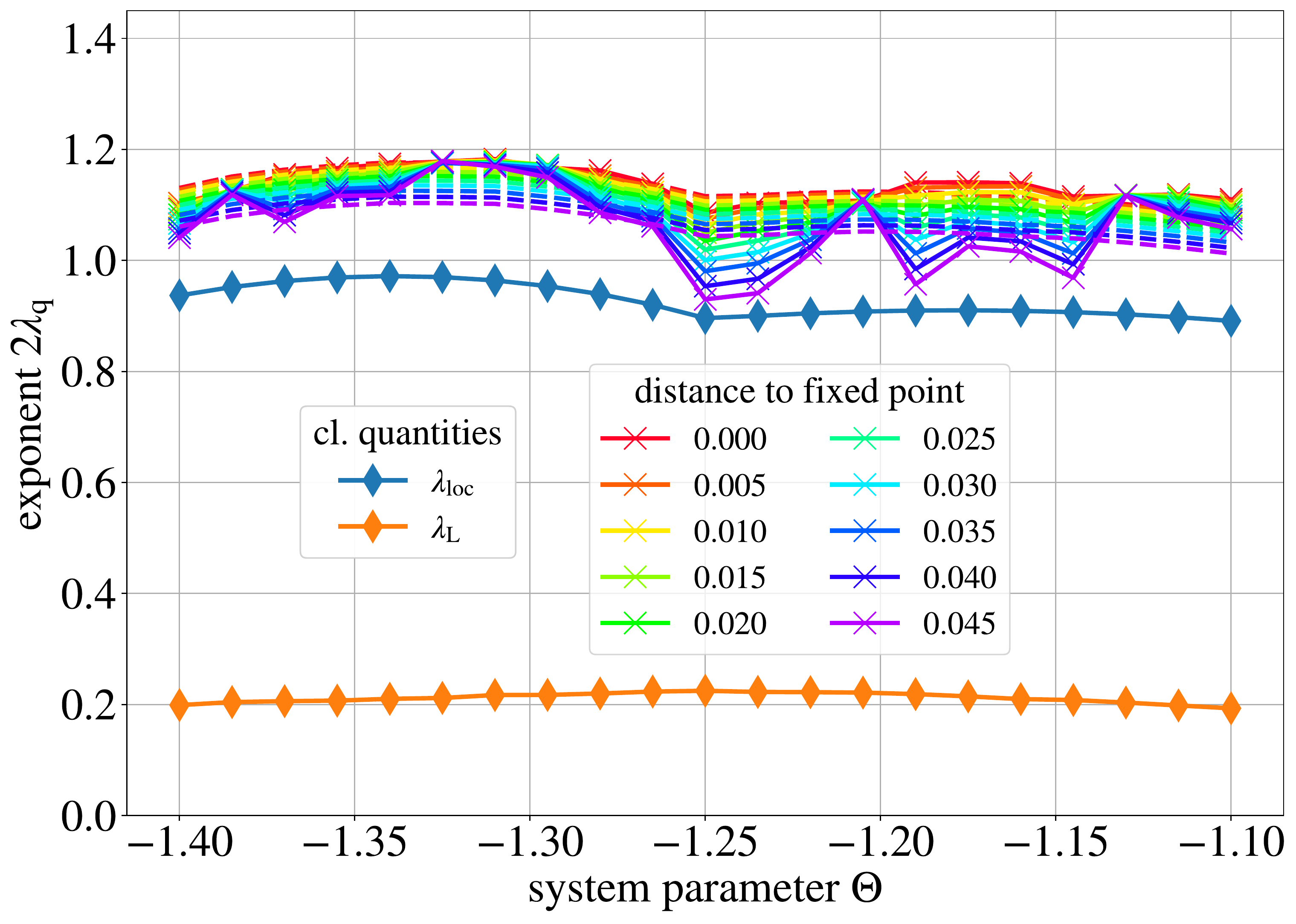}
    \caption{Quantum Lyapunov exponent for states on the energy shell of the fixed point;  upper plot \textbf{a)}: $L=3,~N=100$; lower plot \textbf{b)}: $L=4,~N=40$; color code represents the distance to the fixed point $z^{\rm H}$; dashed lines represent the fit via the hypothesis Eq.~\eqref{hypothesis}.}
    \label{fig:BH_OffFP_exponents}
\end{figure*}

\begin{figure}[h!]
    \centering
    \subfigimg[width=1\linewidth]{\hspace{1.25cm}\textbf{a)}}{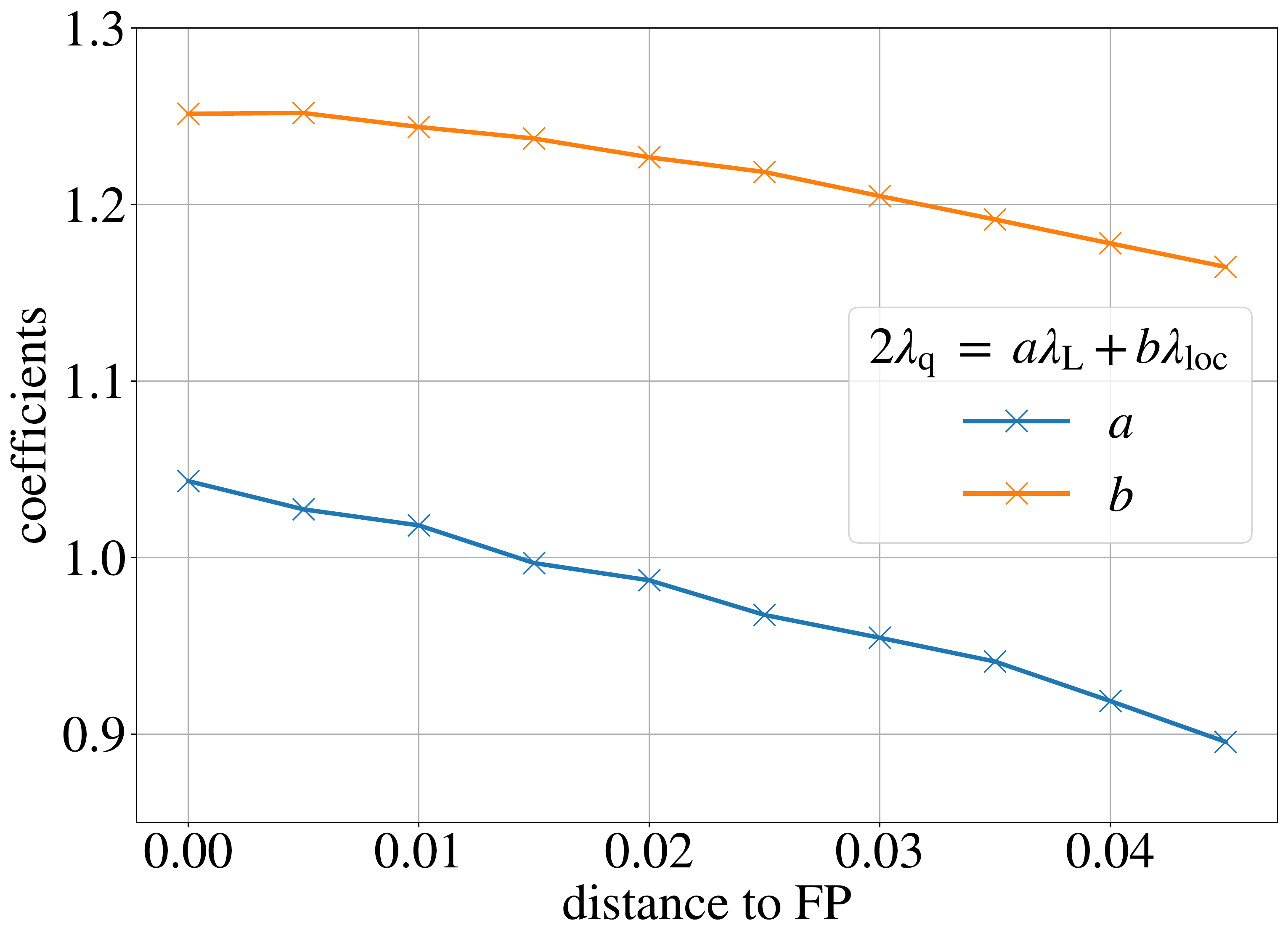}%
    \\
    \subfigimg[width=1\linewidth]{\hspace{1.35cm}\textbf{b)}}{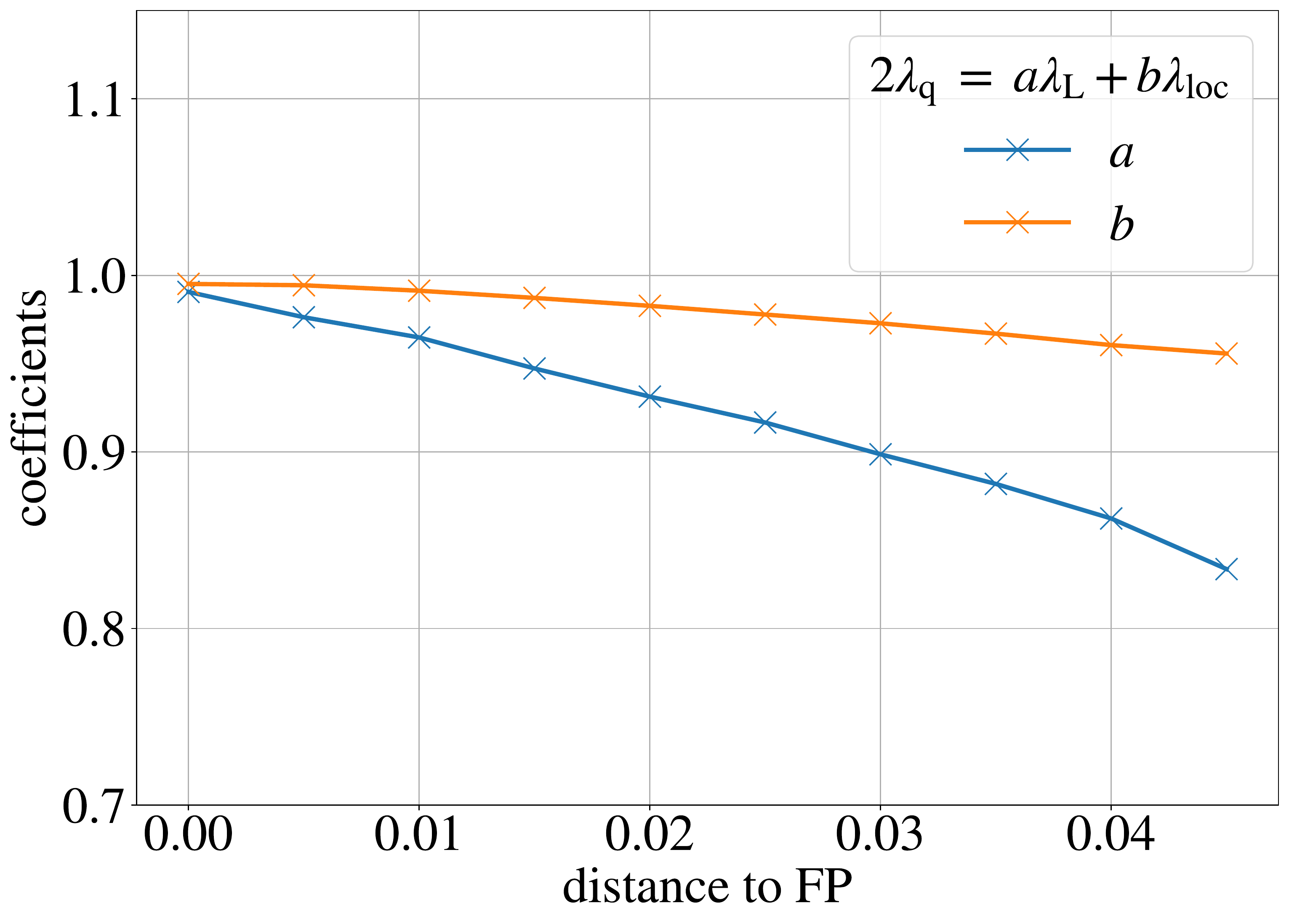}
    \caption{Fitted coefficients $a$ and $b$ from hypothesis Eq.~\eqref{hypothesis};  upper plot \textbf{a)}: $L=3,~N=100$; lower plot \textbf{b)}: $L=4,~N=40$.}
    \label{fig:BH_OffFP_fitCoefficients}
\end{figure}

The system of two interacting large spins from the previous chapter has the clear advantage of having a few enough degrees of freedom so that a very detailed study of the edge of chaos can be carried out, with the corresponding detailed analysis of the OTOC and the hypothesis $2\lambda_{\text{q}}=a\lambda_{\text{L}}+b\lambda_{\text{loc}}$. The expected universality of this conjecture, that arises from generic mechanisms responsible for the emergence of chaotic layers around hyperbolic fixed points should then be checked in other Hamiltonian systems displaying an edge of chaos. We therefore turn now to a system of $N$ spinless Bosons localized in $L$ wells in a ring topology ($L+1\equiv 1)$, which we describe by a (dimensionless) Bose-Hubbard Hamiltonian
\begin{align}
    \hat{H} = -J \sum_{j=1}^{L} \big( \hat{a}^{\dagger}_{j+1} \hat{a}_{j} +\hat{a}^{\dagger}_{j} \hat{a}_{j+1} \big) + \frac{g}{2} \sum_{j=1}^{L}  
     \hat{a}^{\dagger}_{j}\hat{a}^{\dagger}_{j}  \hat{a}_{j} \hat{a}_{j}.
\end{align}
It physically describes ultra cold atom gases in an optical lattices. With the experimental progress in ultra-cold atom gases and optical lattices \cite{ Greiner2002,Bloch2008}, this and similar systems allow us to find many experimental phenomena in the many-body world like tunneling \cite{Albiez2005,Lignier2007},  many-body localization \cite{Choi2016} etc.
The role of the effective Planck constant is played by the inverse of the total particle number $\hbar_{\rm{eff}} = 1/N$ \cite{PhysRevLett.121.124101}.
The first term with coefficient $J$ is the next-neighbor hopping from adjacent wells and the second term with $g$ is the on-site interaction between Bosons. We introduce a convenient system parameter $\Theta$ such that we express $J=\cos(\Theta)$ and $g= \frac{L}{N} \sin(\Theta)$. 
We calculate the OTOC for number-projected coherent states defined by 
\begin{align}
    |\Vec{\phi}\rangle = \frac{1}{\sqrt{N!}} \Big( \Vec{\phi} \cdot \hat{\Vec{a}}^{\dagger}\Big)^{N} | 0\rangle,
\end{align}
where $\Vec{\phi}$ is a complex vector $\phi_{j} = (q_{j}+ i p_{j})/\sqrt{2}$ which is associated to a point of the classical mean-field phase space with $|| \Vec{\phi}||_{2}=1$. The classical mean-field system ($\hbar_{\text{eff}}\to 0$) is given by replacing the operators to complex numbers in the normal ordered Hamiltonian,
\begin{align}
    \begin{split}
         H(\Vec{q},\Vec{p}) =& -\cos(\Theta) \sum_{j=1}^{L} (q_{j}q_{j+1}+ p_{j}p_{j+1}) 
        \\
        &+\frac{\sin(\Theta) L }{8} \sum_{j=1}^{L}(q_{j}^{2} + p_{j}^{2})^{2}
    \end{split}\,\,\,\text{.}
\end{align}
In contrast to the large spin system described in Sec.~\ref{ch3}, the domain in phase space, here given just by $\mathcal{P}=\mathbb{R}^{2L}$, in which the dynamics takes place is a $2 L -1$ dimensional sphere $S^{2L-1}$ (as the particle number is a conserved quantity, i.e., the mean-field norm $||\Vec{\phi}||_{2} = 1 $ is a constant of motion). There is a trivial $U(1)$ symmetry left which must be considered when searching for fixed points. We compensate the global phase by adding a frequency $\frac{\mu}{2} \sum_{j=1}^{L}(q_{j}^{2} + p_{j}^{2})$ to the Hamiltonian $H$, i.e., we search for fixed points of the flow equations as zeros of the Hamiltonian vector field
\begin{align}
	X^H_{(z;\mu)} &= {(dH)}_{(z;\mu)}^{\#_\omega} \\
	&\hspace{0cm}=\left[\begin{array}{l}\
    \big\{ \!-\cos(\Theta) (p_{j+1}+p_{j-1}) +
    \\
    ~~~~+ \frac{\sin(\Theta)L}{2} (q_{j}^{2} + p_{j}^{2}) p_{j} + \mu p_{j}\!\big\}_{j=1,\ldots,L}\!\\
    \big\{\cos(\Theta) (q_{j+1}+q_{j-1}) +
    \\
    ~~~~- \frac{\sin(\Theta)L}{2} (q_{j}^{2} + p_{j}^{2}) q_{j} - \mu q_{j}\big\}_{j=1,\ldots,L}
	\end{array}\right]
	\stackrel{\text{!}}{=} 0
    \nonumber
\end{align}
for a phase space point $z = (q_{1},\ldots,q_{L},p_{1},\ldots,p_{L})$ and a frequency $\mu$. Thus we extend the notion for fixed point for classical mean-field solution having a trivial time-evolution~\cite{EILBECK1985318}.  
We focus on the homogeneous fixed point~$z^{\rm H}$ given by 
\begin{align*}
    q_{j}^{\text{H}}&= \sqrt{\frac{2}{L}},~ p_{j}^{\text{H}}=0,~ \phi_{j}^{\rm H}=\frac{1}{\sqrt{L}}
    \\ 
    \mu^{\text{H}} &=\sin(\Theta) - 2\cos(\Theta).
\end{align*}
A straightforward  stability analysis shows that the fixed point~$z^{\rm H}$ is unstable in the parameter region $-\frac{\pi}{2}<\Theta < \arctan (-1+\cos(\frac{2\pi}{L}))$. 
We want to study the hypothesis~Eq.~\eqref{hypothesis} for the Bose Hubbard rings $L=3, ~N=100$ and $L=4,~N=40$. The homogeneous fixed point is unstable in the parameter region $\Theta\in [-1.4, -1.1]$ for both systems,  as well as embedded inside a chaotic layer of the phase space ($\lambda_{\text{L}}>0$) (see red data points in Fig.~\ref{fig:BH_FP_exponents}). We skip the Poincaré surface of section analysis, as this does not yield a convenient visualization in the high dimensional phase space compared to Fig.~\ref{PhasespacePlots}.
The OTOC we calculate for the Bose-Hubbard system 
\begin{align*}
   C^{n_{1}n_{1}} (t)= \langle \Vec{\phi}^{\, \text{H}} | \big|\big|[\hat{n}_{1}(t),\hat{n}_{1}]\big|\big|^{2} | \Vec{\phi}^{\, \text{H}} \rangle 
\end{align*}
is the squared commutator of the occupations at the first site of the ring  $\hat{n}_{1} = \hat{a}^{\dagger}_{1}\hat{a}_{1}$ at times $t$ and $0$. This kind of OTOC is studied in a related integrable Bose-Hubbard system~\cite{JD5_2} and the quantum Lyapunov exponent is given by the local instability exponent of the fixed point considered there.  
We extract from Fig.~\ref{fig:BH_FP_exponents} that the magnitude of the local instability $\lambda_{\text{loc}}$ is three to four times bigger than the Lyapunov exponent $\lambda_{\text{L}}$ for the fixed point. Thus, we find an optimal exponential fit using Eq.~\eqref{Definition_fitfunction} within the time-window~$[t_{0},t_{\rm E}]$ defined by the local instability of the homogeneous fixed point, i.e.,~$t_0= 1/\lambda_{\text{loc}}$ and~$t_{\rm E} = \ln(N) /\lambda_{\text{loc}}$. 
Our key observation in Fig.~\ref{fig:BH_FP_exponents} is that the quantum Lyapunov exponent~$2\lambda_{\text{q}}$ is roughly the sum of the local instability and the Lyapunov exponent (plots of the OTOC are included in Fig.~\ref{fig:BH_plot_OTOCs}). 
The exact values of the coefficients indicate, however, that they depend on the localization of the initial wave packet and therefore on~$\hbar_{\text{eff}}$ (width of the Wigner-function scales with~$\sqrt{\hbar_{\text{eff}}}$~\cite{schleich2011quantum}). For $L=3$ the coefficient of $\lambda_{\text{loc}}$ is greater than for $L=4$, as the corresponding $\hbar_{\text{eff}}$ is~$1/100$ and~$1/40$ respectively. 
We check how this dependency of $\lambda_{\rm q}$ on the Lyapunov exponent $\lambda_{\rm L}$ and the instability exponent $\lambda_{\text{loc}}$, when we move away from the fixed point. To this end, we fix the classical energy of the homogeneous fixed point, the $q_{1}$-coordinate in the range $[0,q_{1}^{\text{H}}]$ and numerically find the other coordinates. 
The procedure for the Bose-Hubbard system differs from the large spin system in  Sec.~\ref{ch3}, because we can not rely on a line (chosen from the Poincaré plots in Fig.~\ref{PhasespacePlots}) due to the higher dimensional phase space. 
The numerical result of the coordinate search is shown for exemplary values $L=4$, $N=40$ and $\Theta=-1.1$ in Fig.~\ref{fig:BH_coordinates}, where we target distances (measured from $z^{\rm H}$) between $0$ and $0.04$. 
We repeat this step for each $\Theta$-value and for each ring ($L=3,4$). 
With these sets of phase space points, we calculate the OTOCs, carry out the fitting procedure and plot the exponents in Fig. \ref{fig:BH_OffFP_exponents}, where the distance to the fixed point is encoded in the color gradient. The plots of the OTOCs are displayed in Fig. \ref{fig:BH_plot_OTOCs} with the same color code.
The exponential growth rate has the tendency to decrease with the distance to the fixed point, but for several system parameters $\Theta$ we observe a stagnant exponent. This two-part behavior (a decreasing or stagnant OTOC) is directly visible in the OTOC plots in Fig. \ref{fig:BH_plot_OTOCs} for $L=3$ in panel \textbf{a)} versus \textbf{b)}. 
We interpret the different behavior of exponential growth as a complex dependency on the phase space structure. 
For the stagnant exponents, our explanation is that the actual point is not only close to the fixed point but additionally lying on the un-/stable  manifold emerging from the fixed point. Lying on this manifold and being close enough leads the OTOC to be dominated by the fixed point $z^{\rm H}$ and results in stagnant exponents we observe in Fig. \ref{fig:BH_OffFP_exponents}.
In contrast, if the phase space point is not directly related to the fixed point via an un-/stable manifold, there is only the overlap of the wave packet with the fixed point, which is crucial for the exponential growth rate with the local stability exponent $\lambda_{\rm loc}$. Hence, we see a decrease of the exponent because of the hierarchy $\lambda_{\rm L} < \lambda_{\rm loc}$, when we increase the distance to the fixed point and therefore diminish the overlap. 
Our argument is supported by the previous Sec. \ref{ch3}, as we choose the phase space points from the Poincar\'e surface of section in such a way that we are clearly not on the un-/stable manifold.
Contrary to the clean situation displayed in Fig.~\ref{fig:BH_FP_exponents}. The two different behaviors, shown in Fig.~\ref{fig:BH_OffFP_exponents} when moving away from the fixed point $z^{\text{H}}$, are clearly not captured completely by our hypothesis Eq.~\eqref{hypothesis}. We still can find $\Theta$-independent variables~$a$ and~$b$ such that~$a\lambda_{\text{L}} +b \lambda_{\text{loc}}$ fits the quantum Lyapunov exponent $2\lambda_{\rm q}$ (dashed lines in Fig. \ref{fig:BH_OffFP_exponents} display the fit via the coefficients~$a$ and $b$), but with decreasing validity of the fitting.
Further, we observe a general decrease of both coefficients of $a$ and $b$ displayed in Fig. \ref{fig:BH_OffFP_fitCoefficients} versus the distance to the fixed point. Nevertheless, we still are close to 2 for the sum $a+b$ supporting the findings in the Spin systems in Tab. \ref{Table_results} and trace the decreasing sum $a+b$ back to the issue with stagnating exponents for several system parameters.  

We believe that our results for Bose-Hubbard system, however, does not imply that hypothesis~Eq.\eqref{hypothesis} is falsified. On the contrary, our conclusion is that that we need to ensure not to remain on fixed point's manifolds. Otherwise we are still directly connected to the fixed point via the classical time evolution. 

\section{Conclusion}
\label{ch4}

We have addressed the behavior of quantum scrambling right at the edge of chaos. To begin with, the link between these two concepts is possible for systems admitting a semiclassical regime, and thus we focused on coupled spins, with classical limit and semiclassical regime given by the limit $s \gg 1$ of large total spin, and Bose-Hubbard chains with classical (mean-field) limit given by large occupations $N\gg1$. The classical input in our study requires careful control of the transition to chaos near fixed points of the dynamics, and we use on both cases the interaction strength as control parameter.
We have computed Poincaré surfaces of section and used these as a guide to identify ergodic regions of phase space that can be classified by their maximum Lyapunov exponent.

Once the possibility of applying semiclassical methods is guaranteed, fast quantum scrambling is characterized by the semiclassical approximation to  Out-of-Time-Ordered Correlators, thus predicting an exponential growth. While previous works focus on the extreme scenarios of scrambling due to local critical dynamics and fully chaotic motion, we attempt here for a simple characterization of their interplay. Following the KAM theorem of classical mechanics, the seeds of chaotic motion emerge precisely from the phase space regions near hyperbolic fixed points, and therefore these regions, with intricate dynamics, display such interplay. We have carried out large scale numerical simulations that show how the exponential growth of the OTOC can be best described in terms of classical quantities by using a linear combination of the Lyapunov exponent and the local stability exponent of the fixed point instead of by just either of the two.

For classical states further away from the fixed point the dependence on the Lyapunov exponent $\lambda_{\rm L}$ becomes more pronounced compared to the fixed point stability $\lambda_{\text{loc}}$, though for the numerically determined parameters $a,~b$ in the combination $a \lambda_{\rm L} + b \lambda_{\text{loc}}$ their sum stays approximately constant: $a+b\approx2$ for all classical states.
Comparing this with the two extreme cases where a system either only possesses a hyperbolic fixed-point that can give rise to the instability identified with the OTOC growth or only has a single fully chaotic region in phase space so that only the Lyapunov exponent is the relevant classical quantity we see the curious similarity between the prefactors 2 of these classical quantities and the sum of the individual prefactors in our case for a mixed phase space.
One possible interpretation would be that the classical instability reflected in the OTOC is constant and only distributed differently between the two classical quantities depending on the classical state.
Our work seeks to generalize investigations of the quantum to classical correspondence as captured by OTOCs to the situation with mixed phase space, whereas previous investigations highlighted either purely chaotic \cite{Maldacena,LyapunovQMSpinChains} or regular classical systems with hyperbolic fixed point \cite{InvertedOsc,LyapForRegularClassLimit}.

The systems studied by us exemplify a general phenomenon of the dependence of the quantum Lyapunov exponent on the classical quantities $\lambda_{\rm loc},~\lambda_{\rm L}$ and one could extend our methodology to any further system in which both a hyperbolic fixed point as well as chaotic dynamics are present.
Such a system can always be constructed by starting from any integrable system with a hyperbolic fixed point e.g. by adding a time dependent perturbation to the Hamiltonian.

\subsection{Acknowledgments}
FM, DW, TG and JDU thank the German Research Foundation (DFG) for funding within the Dreiburg collaboration through Projects No. Gu431/9-1 and  Ri681/14-1.
MS acknowledges funding by the Studien\-stiftung des Deutschen Volkes. The authors would like to thank Dominik Hahn and Benjamin Geiger for useful discussions and for making available to us their unpublished work  \cite{HahnThesis}. Furthermore we would like to thank Klaus Richter for helpful discussions on related topics.

\bibliography{references.bib}

\section{Appendix}

\subsection{Derivation of Poisson Bracket Relation with Stability Matrix}
\label{appendix1}
For a Hamiltonian System $((\mathcal{P},\omega),H)$ we express the stability matrix in terms of a Poisson bracket by considering the two functions $f=z^\mu\circ\phi^H, g=z^\nu$.
Here $z^\mu, z^\nu$ are arbitrary components of the symplectic chart $z:\mathcal{P}\rightarrow \mathbb{R}^{2f}$ and in order to keep the language precise for the following explicit calculation we denote by $x$ the point $x\in\mathcal{P}$ which is implicitly identified with its chart representation $x=z(x)$ in the main text.
We use square brackets around an entity to denote its matrix representation with respect to a basis, e.g. $[\omega]$ is the usual symplectic matrix when the basis is the coordinate basis of a symplectic chart, while $\omega$ alone is just the abstract symplectic form on $\mathcal{P}$ before a basis is chosen.
Since we only make use of this usual basis induced by canonical coordinates, we do not explicitly denote it next to the square brackets.
Furthermore, indexed expressions inside square brackets denote the matrix obtained by interpreting the first index as a row and the second one as a column index, while indices outside of the brackets are just the indices of an object already viewed as a matrix.
Now we make use of the definition Eq.~\eqref{definition_Poisson} and compute
\begin{align}
\label{poisson_bracket_stability_matrix}
	&\{z^\mu\circ \phi_t,z^\nu\}_{(x)} = 
	\omega_{(x)}\left(X^{z^\mu \circ \phi_t}_{(x)},X^{z^\nu}_{(x)}\right)=\\
	&=\left(\left(d(z^\mu\circ \phi_t)_{(x)}\right)^{\#_\omega}\right)^{\mathfrak{b}_\omega}\left((dz^\nu_{(x)})^{\#_\omega}\right)=\nonumber\\
	& = dz^\mu_{(\phi_t(x))}\circ {d\phi_t}_{(x)}\left(\omega^{\alpha\beta}_{(x)}(dz^\nu_{(x)})_\alpha \frac{\partial}{\partial z^\beta}_{(x)}\right)=\nonumber\\
	& = \omega^{\alpha\beta}_{(x)} \delta^\nu_\alpha dz^\mu_{(\phi_t(x))}\left({d\phi_t}_{(x)}\left(\frac{\partial}{\partial z^\beta}_{(x)}\right)\right)\,\,\,\text{.}\nonumber
\end{align}
Here the inverse operation to $\#_\omega$ is denoted with $\mathfrak{b}_\omega$.
We continue writing the linear map ${d\phi_t}_{(x)} : \mathrm{T}_x \mathcal{P} \rightarrow \mathrm{T}_{\phi_t(x)}\mathcal{P}$ with respect to a basis of coordinate one-forms $dz^\gamma_{(x)}$ in the domain and a vector basis $\partial/\partial z^\eta_{(\phi_t(x))}$ in the target space and insert it into the previous expression:
\begin{align}
	& \omega^{\alpha\beta}_{(x)} \delta^\nu_\alpha dz^\mu_{(\phi_t(x))}\left({d\phi_t}_{(x)}\left(\frac{\partial}{\partial z^\beta}_{(x)}\right)\right)=\\
	& = \omega^{\alpha\beta}_{(x)} \delta^\nu_\alpha dz^\mu_{(\phi_t(x))}\cdot\nonumber\\
	&\cdot \left(
		[{d\phi_t}_{(x)}]^{\eta}_{\,\,\,\gamma} dz^{\gamma}_{(x)} 
		\left(
			\frac{\partial}{\partial z^\beta}_{(x)}
		\right)
		\otimes \frac{\partial}{\partial z^\eta}_{(\phi_t(x))}
	\right) =\nonumber\\
	& = \omega^{\nu\beta}_{(x)} [{d\phi_t}_{(x)}]^\eta_{\,\,\,\gamma} \delta^\gamma_\beta 
	dz^\mu_{(\phi_t(x))}\left( \frac{\partial}{\partial z^\eta}_{(\phi_t(x))}\right)=\nonumber\\
	& = \omega^{\nu\gamma}_{(x)} [{d\phi_t}_{(x)}]^\eta_{\,\,\,\gamma} \delta^\mu_\eta= \omega^{\nu\gamma}_{(x)} [{d\phi_t}_{(x)}]^\mu_{\,\,\,\gamma}\,\,\,{\text{.}}\nonumber
\end{align}
If we write the whole expression as a matrix with entries $[\{z^\mu\circ \phi_t,z^\nu\}_{(x)}]$ in the $\mu$-th row and $\nu$-th column, we finally obtain
\begin{align}
\label{poisson_bracket_stability_final}
	&[\{z^\mu\circ \phi_t,z^\nu\}_{(x)}]=\\
	&= {[\omega_{(x)}]^{-1}}^T[{d\phi_t}_{(x)}]=[\omega][{d\phi_t}_{(x)}]\,\,\,\text{.}\nonumber
\end{align}
Hence we can now express the matrix representation of the linearized flow with respect to the coordinate basis corresponding to a chart $z$ by above Poisson bracket.

\subsection{Energy Drift of Numerics for Phase Space Plots}
\label{appendix2}
\begin{figure*}
    \centering
	\includegraphics[width=0.49\textwidth]{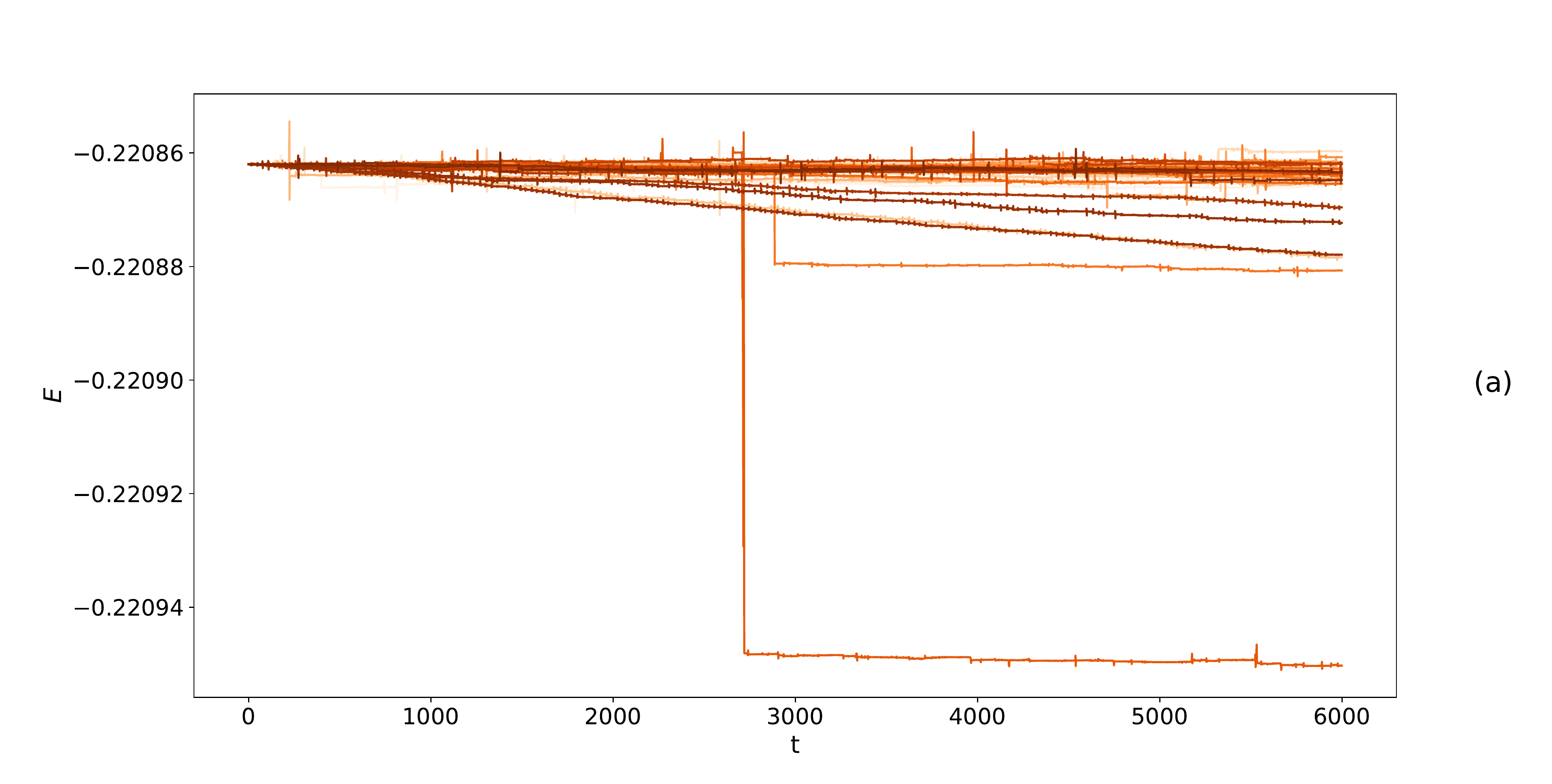}%
	\includegraphics[width=0.49\textwidth]{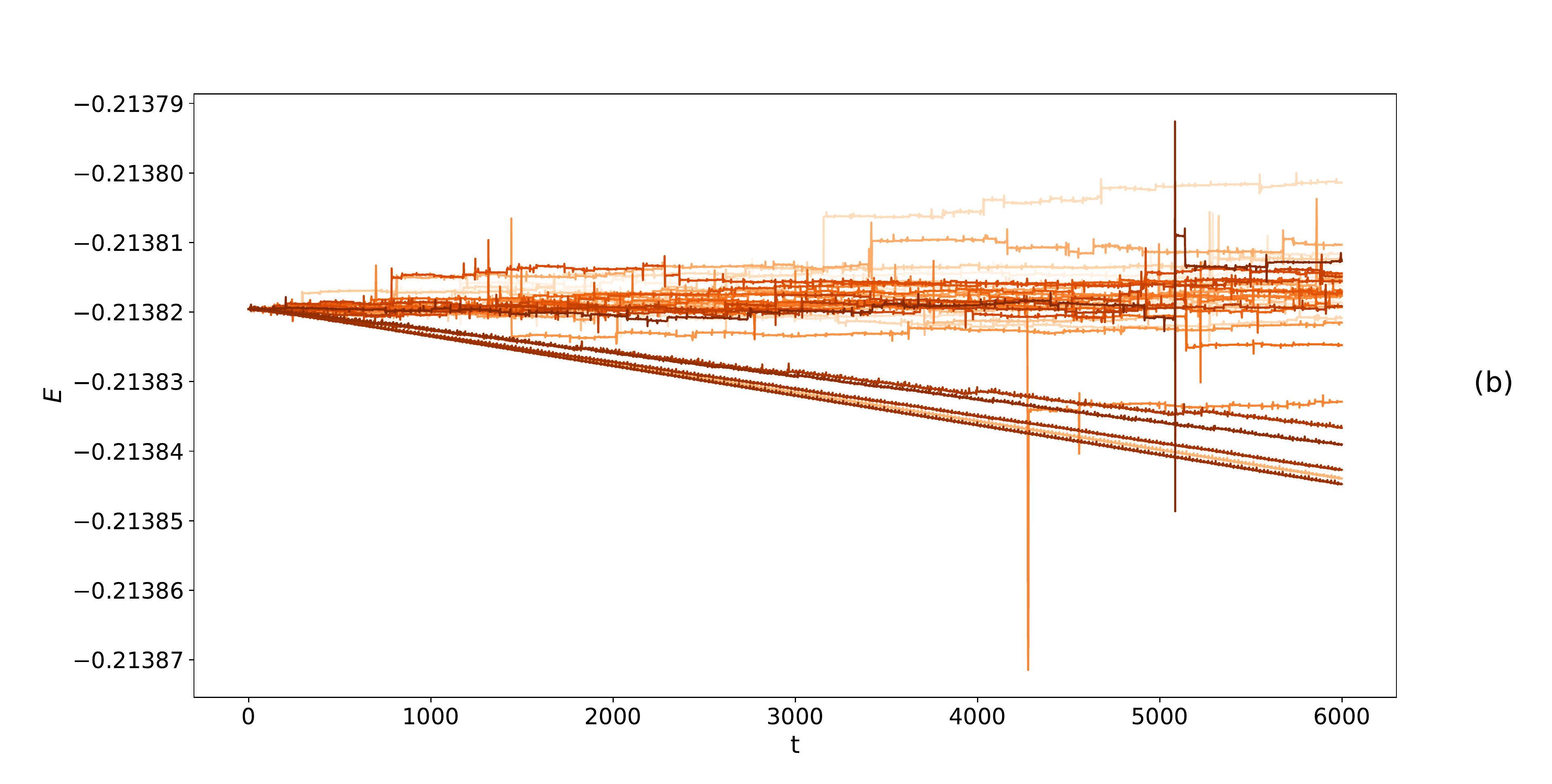}
	\\
	\includegraphics[width=0.49\textwidth]{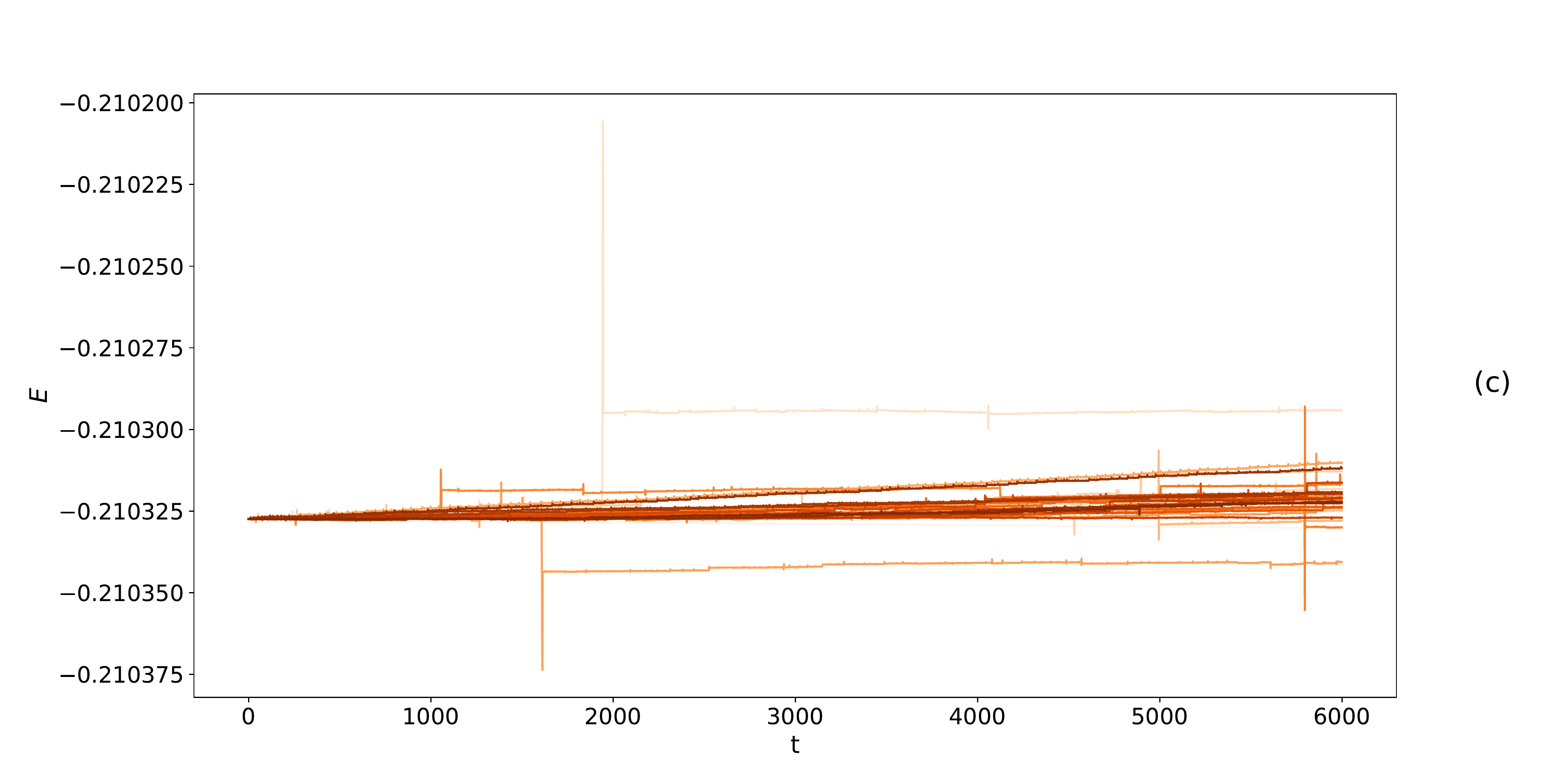}%
	\includegraphics[width=0.49\textwidth]{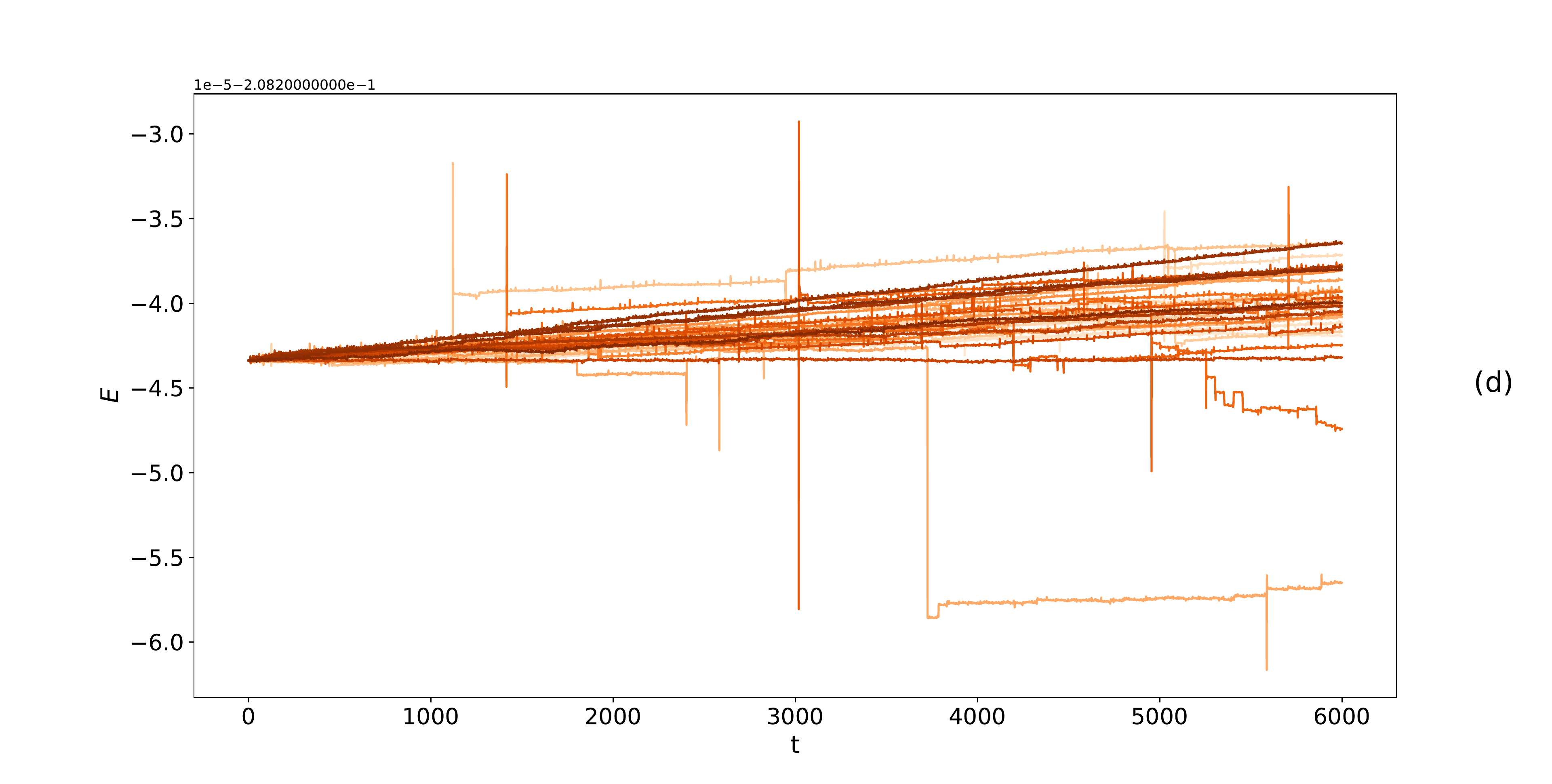}
    \caption{Energy drifts of numerically computed trajectories on same energy shell for various initial conditions on the same energy shell distinguished by color shading. In plot (d) a global factor $10^{-5}$ has to be kept in mind as written in the top left corner of the respective plot. The labeling (a) to (d) corresponds to the plots in Fig.~\ref{PhasespacePlots}. At $t=0$ all plots start at the desired energies listed in Sec.~\ref{ch3}.}
    \label{EnergyDrifts}
\end{figure*}
Here we present the energy drifts of the individual trajectories comprising the energy sections depicted in Sec.~\ref{ch3} (see Fig.~\ref{PhasespacePlots}).
While the Hamiltonian of the spin system is time independent and hence a constant of motion, unavoidable numerical errors result in a slight deviation from this theoretical fact so that a numerically computed solution of the equations of motion does not consist only of points of the exact same energy.
In the plots in Fig.~\ref{EnergyDrifts}, we show the energy of each point on each trajectory that is plotted as part of an energy section in Fig. \ref{PhasespacePlots}.
Each color shade represents points on the same trajectory.
One sees that energy conservation is obeyed sufficiently well by the numerics with only individual points from some of the time intervals of 6000 time steps having instantaneous large deviations from the average.
Even those anomalous deviations stay within the same order of magnitude as the average though.

\end{document}